\documentclass[titlepage,pallis,twoside]{wpallis}
\usepackage{fancyh}
\usepackage[l]{floatflt}
\usepackage{epsfig}
\usepackage{amssymb}
\usepackage{latexsym}
\usepackage{times}
\usepackage{slashed,cite}
\usepackage{upgreek}
\numberwithin{equation}{section}

\usepackage{amsmath}
\usepackage{amsfonts}
\usepackage{amsbsy}
\usepackage{amscd}
\usepackage{bbm}

\newcommand{\bal}{\begin{align}}
\newcommand{\eal}{\end{align}}
\newcommand{\beqs}{\begin{subequations}}
\newcommand{\eeqs}{\end{subequations}}
\newcommand{\ec}{\end{center}}
\newcommand{\bec}{\begin{center}}

\newcommand{\eem}{\end{matrix}}
\newcommand{\bem}{\begin{matrix}}
\newcommand{\eeq}{\end{equation}}
\newcommand{\beq}{\begin{equation}}
\newcommand{\ba}{\begin{array}}
\newcommand{\ea}{\end{array}}
\newcommand{\bea}{\begin{eqnarray}}
\newcommand{\eea}{\end{eqnarray}}
\newcommand{\baq}{\begin{eqnarray}}
\newcommand{\eaq}{\end{eqnarray}}

\newcommand\eqs[2]{Eqs.~(\ref{#1}) and (\ref{#2})}

\newcommand\eqss[3]{Eqs.~(\ref{#1}), (\ref{#2}) and (\ref{#3})}
\newcommand\eqsss[4]{Eqs.~(\ref{#1}), (\ref{#2}), (\ref{#3}) and (\ref{#4})}

\newcommand{\ftn}{\footnotesize}

\newcommand{\GeV}{{\mbox{\rm GeV}}}

\newcommand{\sFig}[2]{Fig.~\ref{#1}-${\sf ({#2})}$}
\newcommand{\sFref}[2]{Fig.~\ref{#1}-{\ftn\sf ({#2})}}
\newcommand{\sEref}[2]{Eq.~(\ref{#1}{\ftn\sf {#2}})}

\newcommand{\etal}{{\it et al.\/}}

\def\to{\rightarrow}

\def\lf{\left(}
\def\rg{\right)}
\newcommand\vev[1]{\langle {#1} \rangle}

\newcommand{\Vhi}{\ensuremath{\widehat V_{\rm IG}}}
\newcommand{\Hhi}{\ensuremath{\widehat H_{\rm IG}}}
\newcommand{\Ohi}{\ensuremath{\Omega}}

\newcommand{\Khi}{\ensuremath{K}}
\newcommand{\Whi}{\ensuremath{W}}
\newcommand{\Vhio}{\ensuremath{\widehat V_{\rm IG0}}}

\newcommand{\mP}{\ensuremath{m_{\rm P}}}

\newcommand{\Qef}{\ensuremath{\Lambda_{\rm UV}}}

\def\openone{\leavevmode\hbox{\small1\kern-3.8pt\normalsize1}}

\newcommand{\ft}{\ensuremath{f_\Phi}}
\newcommand{\kx}{\ensuremath{k_S}}

\newcommand{\Fcr}{\ensuremath{\Omega_{\rm H}}}
\newcommand{\fr}{\ensuremath{f_{\cal R}}}
\newcommand{\fk}{\ensuremath{\Omega_{\rm H}}}

\newcommand{\fp}{\ensuremath{f_\Phi}}
\newcommand{\fsp}{\ensuremath{f_{S\Phi}}}
\newcommand{\Fk}{\ensuremath{\Omega_{\rm K}}}
\newcommand{\ks}{\ensuremath{k_S}}
\newcommand{\ksp}{\ensuremath{k_{S\Phi}}}
\newcommand{\kpp}{\ensuremath{k_{\Phi}}}

\newcommand{\ck}{\ensuremath{c_{\cal R}}}

\newcommand{\msn}{\ensuremath{\what m_{\rm \dph}}}

\newcommand{\ns}{\ensuremath{n_{\rm s}}}
\newcommand{\as}{\ensuremath{a_{\rm s}}}
\newcommand{\As}{\ensuremath{A_{\rm s}}}
\newcommand{\rcc}{\ensuremath{\mathcal{R}}}
\newcommand{\rce}{\ensuremath{\widehat{\mathcal{R}}}}
\newcommand{\Ve}{\ensuremath{\widehat{V}}}
\newcommand{\He}{\ensuremath{{\what H}}}
\newcommand{\Ne}{\ensuremath{{\what N}}}
\newcommand{\Ns}{\ensuremath{{\what N_\star}}}

\newcommand{\dphi}{\ensuremath{\what{\delta\phi}}}
\newcommand{\dph}{\ensuremath{\delta\phi}}

\newcommand{\what}{\ensuremath{\widehat}}

\def\aal{{\bar\alpha}}
\def\bbet{{\bar\beta}}
\def\al{{\alpha}}

\def\bt{{\beta}}

\def\K{{\widehat{K}}}

\def\th{{\theta}}

\newcommand{\Trh}{\ensuremath{T_{\rm rh}}}
\newcommand{\sg}{\ensuremath{\phi}}

\newcommand{\sgx}{\ensuremath{\phi_\star}}
\newcommand{\sgf}{\ensuremath{\phi_{\rm f}}}
\newcommand{\xsg}{\ensuremath{x_{\phi}}}

\newcommand{\ld}{\ensuremath{\lambda}}

\newcommand{\se}{\ensuremath{\widehat \phi}}
\newcommand{\sex}{\ensuremath{\widehat{\phi}_\star}}

\newcommand{\geu}{\ensuremath{\widehat g}}

\def\trns{transplanckian}
\def\Ka{K\"{a}hler potential}
\def\Km{K\"{a}hler manifold}

\def\sub{subplanckian}

\def\FHI{IG inflation~}

\newcommand{\diag}{\ensuremath{{\sf diag}}}

\renewcommand{\arg}{\ensuremath{{\small\sf arg}}}

\newcommand{\bicep}{{\scshape Bicep2}}

\begin{document}

\thispagestyle{empty}

\title[]{\boldmath\Large\bfseries\scshape
Induced-Gravity Inflation in \\ no-Scale Supergravity and Beyond}

\author{\large\bfseries\scshape C. Pallis}
\address[] {\sl Departament de F\'isica Te\`orica and IFIC,\\
Universitat de Val\`encia-CSIC, \\ E-46100 Burjassot, SPAIN}

\begin{abstract}{{\bfseries\scshape Abstract} \\
\par Supersymmetric versions of induced-gravity inflation are
formulated within Supergravity (SUGRA) employing two gauge singlet
chiral superfields. The proposed superpotential is uniquely
determined by applying a continuous $R$ and a discrete
$\mathbb{Z}_n$ symmetry. We select two types of logarithmic \Ka s,
one associated with a no-scale-type $SU(2,1)/SU(2)\times
U(1)_R\times\mathbb{Z}_n$ \Km\ and one more generic. In both
cases, imposing a lower bound on the parameter $\ck$  involved in
the coupling between the inflaton and the Ricci scalar curvature
-- e.g. $\ck\gtrsim 76, 105, 310$ for $n=2,3$ and $6$ respectively
--, inflation can be attained even for \sub\ values of the
inflaton while the corresponding effective theory respects the
perturbative unitarity. In the case of no-scale SUGRA we show
that, for every $n$, the inflationary observables remain unchanged
and in agreement with the current data while the inflaton mass is
predicted to be $3\cdot10^{13}~\GeV$. Beyond no-scale SUGRA the
inflationary observables depend mildly on $n$ and crucially on the
coefficient involved in the fourth order term of the \Ka\ which
mixes the inflaton with the accompanying non-inflaton field. }
\\ \\
{\ftn \sf Keywords: Cosmology, Supersymmetric models, Supergravity, Modified Gravity};\\
{\ftn \sf PACS codes: 98.80.Cq, 11.30.Qc, 12.60.Jv, 04.65.+e,
04.50.Kd}\\ \\ \publishedin{{\sl  J. Cosmol. Astropart. Phys.}
{\bf 08}, {057} (2014)}
\end{abstract} \maketitle



\setcounter{page}{1} \pagestyle{fancyplain}

\rhead[\fancyplain{}{ \bf \thepage}]{\fancyplain{}{\sl IG
Inflation in no-Scale SUGRA \& Beyond}} \lhead[\fancyplain{}{ \sl
\leftmark}]{\fancyplain{}{\bf \thepage}} \cfoot{}

\tableofcontents\vskip-1.3cm\noindent\rule\textwidth{.4pt}

\section{Introduction}\label{intro} %

The announcement of the recent PLANCK results \cite{wmap,plin}
fuelled increasing interest in inflationary models implemented
thanks to a strong enough non-minimal coupling between the
inflaton field, $\phi$, and the Ricci scalar curvature, $\rcc$.
Indeed, these models predict \cite{plin, defelice13} a (scalar)
spectral index $\ns$, tantalizingly close to the value favored by
observational data. The existing non-minimally coupled to Gravity
inflationary models can be classified into two categories
depending whether the non-minimal coupling to $\rcc$ is added into
the conventional one, $\mP^2\rcc/2$ -- where $\mP = 2.44\cdot
10^{18}~\GeV$ is the reduced Planck scale -- or it replaces the
latter. In the first case the \emph{vacuum expectation value}
({\ftn\sf v.e.v}) of the inflaton after inflation assumes
sufficiently low values after inflation, such that a transition to
Einstein gravity at low energy to be guarantied. In the second
case, however, the term $\mP^2\rcc/2$ is dynamically generated via
the v.e.v of the inflaton; these models are, thus, named
\cite{induced, higgsflaton} \emph{Induced-Gravity} ({\ftn\sf IG})
inflationary models. Despite the fact that both models of
non-Minimal Inflation are quite similar during inflation and may
be collectively classified into universal ``attractor'' models
\cite{roest}, they exhibit two crucial differences. Namely, in the
second category, {\sf\ftn (i)} the \emph{Einstein frame} ({\ftn\sf
EF}) inflationary potential develops a singularity at $\phi=0$ and
so, inflation is of Starobinsky-type \cite{R2} actually; {\sf\ftn
(ii)}  The \emph{ultaviolet} ({\ftn\sf UV}) cut-off scale
\cite{cutoff,cutof,riotto} of the theory, as it is recently
realized \cite{pallis,gian}, can be identified with $\mP$ and,
thereby, concerns regarding the naturalness of inflation can be
safely eluded. On the other hand, only some \cite{riotto} of the
remaining models of nonminimal inflation can be characterized as
unitarity safe.

In a recent paper \cite{pallis} a \emph{supersymmetric} ({\ftn\sf
SUSY}) version of IG inflation was, for first time, presented
within no-scale \cite{noscale,eno5, eno7} \emph{Supergravity}
({\ftn\sf SUGRA}). A Higgs-like modulus plays there the role of
inflaton, in sharp contrast to \cref{eno5} where the inflaton is
matter-like. For this reason we call in \cref{pallis} the
inflationary model \emph{no-scale modular inflation}. Although any
connection with the no-scale SUSY breaking \cite{noscale, eno9} is
lost in that setting, we show that the model provides a robust
cosmological scenario linking together non-thermal leptogenesis,
neutrino physics and a resolution to the $\mu$ problem of the
\emph{Minimal SUSY SM} ({\ftn\sf MSSM}). Namely, in \cref{pallis},
we employ a \Ka, $K$, corresponding to a $SU(N,1)/ SU(N)\times
U(1)_R\times\mathbb{Z}_2$ symmetric \Km. This symmetry fixes
beautifully the form of $K$ up to an holomorphic function $\fk$
which exclusively depends on the inflaton, $\phi$, and its form
$\fk\sim\phi^2$ is fixed by imposing a $\mathbb{Z}_2$ discrete
symmetry which is also respected by the superpotential $\Whi$.
Moreover, the model possesses a continuous $R$ symmetry, which
reduces to the well-known $R$-parity of MSSM. Thanks to the strong
enough coupling between $\phi$ and $\rcc$, inflation can be
attained even for \sub\ values of $\phi$, contrary to other SUSY
realizations \cite{eno7,linde,zavalos} of the Starobinsky-type
inflation.

Most recently a more generic form of $\fk$ has been proposed
\cite{gian} at the non-SUSY level. In particular, $\fk$ is
specified as $\fk\sim\phi^n$ and it was pointed out that the
resulting IG inflationary models exhibit an attractor behavior
since the inflationary observables and the mass of the inflaton at
the vacuum are independent of the choice of $n$. It would be,
thereby, interesting to investigate if this nice feature insists
also in the SUSY realizations of these models. This aim gives us
the opportunity to generalize our previous analysis \cite{pallis}
and investigate the inflationary predictions independently of the
post-inflationary cosmological evolution. Namely, we here impose
on $\fk$ a discrete $\mathbb{Z}_n$ symmetry with $n\geq2$, and
investigate its possible embedding in the standard Poincar\'e
SUGRA, without invoking the superconformal formulation --
cf.~\cref{rena}. We discriminate two possible embeddings, one
based on a no-scale-type symmetry and one more generic, with the
first of these being much more predictive. Namely, while the
embedding of IG models in generic SUGRA gives adjustable results
as regards the inflationary observables, -- see also \cref{talk}
--, no-scale SUGRA predicts independently of $n$ results identical
to those obtained in the non-SUSY case. Therefore, no-scale SUGRA
consists a natural framework in which such models can be
implemented.

Below, in Sec.~\ref{fhim}, we describe the generic formulation of
IG models within SUGRA. In Sec.~\ref{fhi} we present the basic
ingredients of our IG inflationary models, derive the inflationary
observables and confront them with observations. We also provide a
detailed analysis of the UV behavior of these models in
Sec.~\ref{fhi3}. Our conclusions are summarized in Sec.~\ref{con}.
Throughout the text, the subscript of type $,\chi$ denotes
derivation  \emph{with respect to} ({\ftn\sf w.r.t}) the field
$\chi$ (e.g., $_{,\chi\chi}=\partial^2/\partial\chi^2$) and charge
conjugation is denoted by a star.

\section{Embedding IG Inflation in SUGRA}\label{fhim}

In \Sref{sugra1} we present the basic formulation of a theory
which exhibits non-minimal coupling of scalar fields to $\rcc$
within SUGRA and in \Sref{sugra2} we outline our strategy in
constructing viable models of IG inflation. The general framework
for the analysis of the emerged models is given in \Sref{sugra3}.

\subsection{The General Set-up} \label{sugra1}

Our starting point is the EF action for $N$ gauge singlet scalar
fields $z^\al$ within SUGRA \cite{linde1,nmN} which can be written
as
\beqs \beq\label{Saction1} {\sf S}=\int d^4x \sqrt{-\what{
\mathfrak{g}}}\lf-\frac{1}{2}\mP^2 \rce +K_{\al\bbet}\geu^{\mu\nu}
\partial_\mu z^\al \partial_\nu z^{*\bbet}-\Ve\rg, \eeq
where summation is taken over the scalar fields $z^\al$,
$K_{\al\bbet}={\K_{,z^\al z^{*\bbet}}}$ with
$K^{\bbet\al}K_{\al\bar \gamma}=\delta^\bbet_{\bar \gamma}$,
$\widehat{\mathfrak{g}}$ is the determinant of the EF metric
$\geu_{\mu\nu}$, $\rce$ is the EF Ricci scalar curvature, $\Ve$ is
the EF F--term SUGRA scalar potential which can be extracted once
the superpotential $\Whi$ and the \Ka\ $\Khi$ have been selected,
by applying the standard formula
\beq \Ve=e^{\Khi/\mP^2}\left(K^{\al\bbet}{\rm F}_\al {\rm
F}^*_\bbet-3\frac{\vert
W\vert^2}{\mP^2}\right),\>\>\>\mbox{where}\>\>\> {\rm
F}_\al=W_{,z^\al} +K_{,z^\al}W/\mP^2.\label{Vsugra} \eeq \eeqs
Note that D-term contributions into $\Ve$ do not exist since we
consider gauge singlet $z^\al$'s. By performing a conformal
transformation and adopting a frame function $\Omega$ which is
related to $K$ as follows
\beq-\Omega/3
=e^{-K/3\mP^2}\>\Rightarrow\>K=-3\mP^2\ln\lf-\Omega/3\rg,\label{Omg1}\eeq
we arrive at the following action
\beq {\sf S}=\int d^4x
\sqrt{-\mathfrak{g}}\lf-\frac{\mP^2}{2}\lf-{\Omega\over3}\rg
\rcc+\mP^2\Omega_{\al{\bbet}}\partial_\mu z^\al \partial^\mu
z^{*\bbet}-\Omega {\cal A}_\mu{\cal A}^\mu/\mP^2-V \rg,
\label{Sfinal}\eeq
where $g_{\mu\nu}=-\lf 3/\Omega\rg\geu_{\mu\nu}$ and $V
={\Omega^2\Ve}/{9}$ are the JF metric and potential respectively,
we use the shorthand notation $\Omega_\al=\Omega_{,z^\al}$ and
$\Omega_\aal=\Omega_{,z^{*\aal}}$ and ${\cal A}_\mu$ is the purely
bosonic part of the on-shell value of an auxiliary field given by
\beq {\cal A}_\mu =-i\mP^2\lf \Omega_\al\partial_\mu
z^\al-\Omega_\aal\partial_\mu z^{*\aal}\rg/2\Omega\,.
\label{Acal}\eeq
It is clear from \Eref{Sfinal} that ${\sf S}$ exhibits non-minimal
couplings of the $z^\al$'s to $\rcc$. However, $\Omega$ enters the
kinetic terms of the $z^\al$'s too. In general, $\Omega$ can be
written as \cite{linde1}
\beq -\Omega/3=\Fcr(z^\al)+{\Fcr}^*(z^{*\aal})-\Fk\lf z^\al
z^{*\aal}\rg/3, \label{Omg}\eeq
where $\Fk$ is a dimensionless real function while $\Fcr$ is a
dimensionless, holomorphic function. For $\Fcr>\Fk$, $\Fk$
expresses mainly the kinetic terms of the $z^\al$'s whereas $\Fcr$
represents the non-minimal coupling to gravity -- note that
$\Omega_{\al{\bbet}}$ is independent of $\Fcr$ since $\Omega_{{\rm
H},z^\al z^{*\bbet}}=0$.

To realize the idea of IG, we have to assume that $\fk$ depends on
a Higgs-like modulus, $z^1:=\Phi$ whose the v.e.v generates the
conventional term of the Einstein gravity at the SUSY vacuum, i.e.
\beq
\vev{\fk}+\vev{\fk^*}=1~~\Rightarrow~~\vev{\fk}=1/2\>\>\>\mbox{for}\>\>\>\vev{\Fk}\sim0
\label{ig}\eeq
where we take into account that the phase of $\Phi$, $\arg\Phi$ is
stabilized to zero; we thus get $\vev{\fk}=\vev{\fk^*}$.

In order to get canonical kinetic terms, we need \cite{linde1}
${\cal A}_\mu=0$ and $\Omega_{{\rm K}\al{\bbet}}\simeq0$ or
$\delta_{\al{\bbet}}$. The first condition is attained when the
dynamics of the $z^\al$'s is dominated only by the real moduli
$|z^\al|$. The second condition is satisfied by the choice
\beq \label{Fkdef} \Fk\lf|z^\al|^2 \rg= k_\al
{|z^\al|^2/\mP^2}\,-\, k_{\al\bt}\, {|z^\al|^2|z^\bt|^2/\mP^4}\eeq
with sufficiently small coefficients $k_\al$ and
$k_{\al\bt}\simeq1$. Here we assume that the $z^\al$'s are charged
under a global symmetry, so as mixed terms of the form
$z^\al{z}^*_{\bbet}$ are disallowed. The inclusion of the fourth
order term for the accompanying non-inflaton field, $z^2:=S$ is
obligatory in order to evade \cite{linde1} a tachyonic instability
occurring along this direction during IG inflation. As a
consequence, all the allowed terms are to be considered in the
analysis for consistency. Let us here note that such a consistency
is not observed in the SUGRA incarnations of similar models
\cite{linde1,roest}. On the other hand, if we assume that
\beq
k_1=0\>\>\>\mbox{and}\>\>\>k_{1\al}=0,\>\>\>\forall\,\al=1,...,N-1\label{nsks}\eeq
the emergent \Km\ associated with $K$ can be identified with
$SU(N,1)/ SU(N)\times U(1)_R\times\mathbb{Z}_n$ -- where the
symmetries $U(1)_R$ and $\mathbb{Z}_n$ are specified in
\Sref{sugra2} -- and highly simplifies the realization of IG
inflation. The option in \Eref{nsks} is inspired by the early
models of soft SUSY breaking \cite{noscale} and defines
\cite{eno7} no-scale SUGRA. We below show details of these two
realizations of IG inflation.

\subsection{Modeling IG Inflation in SUGRA}\label{sugra2}

As we anticipated above, the realization of the idea of IG in
SUGRA requires at least two singlet superfields, i.e.,
$z^\al=\Phi,S$; $\Phi$ is a Higgs-like superfield whose the v.e.v
generates $\mP$ and $S$ is an accompanying superfield, whose the
stabilization at the origin assists us to isolate the contribution
of $\Phi$ into $\Ve$, \Eref{Vsugra}. To see how this structure
works, let us below specify the form of $\fk$ and $W$.

Inspired by \cref{gian}, we here determine $\fk$ by postulating
its invariance under the action of a global $\mathbb{Z}_n$
discrete symmetry. Therefore it can be written as
\beq \fk(\Phi)=\ck
\frac{\Phi^n}{\mP^n}+\sum_{k=1}^\infty\lambda_{k}\frac{\Phi^{2kn}}{\mP^{2kn}}
\label{fdef}\eeq
with $k$ being a positive integer. Restricting ourselves to \sub\
values of $\Phi$ and assuming relatively low $\lambda_{k}$'s, we
can say that $\mathbb{Z}_n$ uniquely determines the form of $\fk$.
Confining ourselves to a such situation we ignore henceforth the
$k$-dependent terms in \Eref{fdef}. On the other hand, $W$ has to
be selected so as to achieve the arrangement of \Eref{ig}. The
simplest choice is that used in the models of F-term hybrid
inflation \cite{susyhybrid}. As a consequence $\fk(\Phi)$ has to
be involved also in the superpotential $W$ of our model which has
the form
\beq\label{Whi} W= \ld\mP^2 S\lf \fk-1/2\rg/\ck \eeq
and can be uniquely determined if we impose, besides
$\mathbb{Z}_n$, a nonanomalous $R$ symmetry $U(1)_{R}$ under which
\beq S\ \to\ e^{i\alpha}\,S,\>\>\>\fk\ \to\ \fk,\>\>\>W \to\
e^{i\alpha}\, W.\label{Rsym} \eeq
Indeed, $U(1)_R$ symmetry ensures the linearity of $\Whi$ w.r.t
$S$ which is crucial for the success of our construction. To
verify that $W$ leads to the desired $\vev{\fk}$ we minimize the
SUSY limit, $V_{\rm SUSY}$, of $\Ve$, obtained from the latter,
when $\mP$ tends to infinity. This is
\beqs\beq V_{\rm SUSY}= \ld^2\mP^4\left| \fk- 1/2\right|^2/\ck^2 +
\ld^2\mP^4|S\Omega_{{\rm H},\Phi}|^2/\ck^2, \label{VF}\eeq
where the complex scalar components of $\Phi$ and $S$ are denoted
by the same symbol. From \Eref{VF}, we find that the SUSY vacuum
lies at
\beq \vev{S}=0\>\>\>\mbox{and}\>\>\> \vev{\fk}=1/2,\label{vevs}
\eeq\eeqs
as required by \Eref{ig}. Let us emphasize that soft SUSY breaking
effects explicitly break $U(1)_R$ to a discrete subgroup. Usually
\cite{pallis} combining the latter with the $\mathbb{Z}_2^{\rm f}$
fermion parity, yields the well-known $R$-parity of MSSM, which
guarantees the stability of the lightest SUSY particle and
therefore it provides a well-motivated CDM candidate.

The selected $W$ and $K$ by construction give also rise to a stage
of IG inflation. Indeed, placing $S$ at the origin, the only
surviving term of $\Ve$ in \Eref{Vsugra} is
\beqs\beq \Vhio= e^{K/\mP^2}K^{SS^*}\,
|W_{,S}|^2=\frac{\ld^2\mP^4|2\fk-1|^2}{4\ck^2\fsp\fr^2}\>\>\>\mbox{since}\>\>\>e^{K/\mP^2}=\frac{1}{\fr^3}
\>\>\>\mbox{and}\>\>\>K^{SS^*}=\frac{\fr}{\fsp},\label{Vhig}\eeq
where the functions $\fr$ and $\fsp$ are computed along the
inflationary track, i.e.,
\beq \label{frsp}
\fr=-\Ohi/3\>\>\>\mbox{and}\>\>\>\fsp=\mP^2\Omega_{,SS^*}
\>\>\>\mbox{for}\>\>\>S=\arg\Phi=0.\eeq\eeqs
Given that $\fsp\ll\fr\simeq2\fk$ with $\ck\gg1$, an inflationary
plateau emerges since the resulting $\Vhio$ in \Eref{Vhig} is
almost constant. Therefore, $\Phi$ involved in the definition of
$\fk$, \Eref{fdef}, arises naturally as an inflaton candidate.
Note that the non-vanishing values of $\Phi$ during IG inflation
break spontaneously the imposed $\mathbb{Z}_n$; no domain walls
are thus produced due to the spontaneous breaking of
$\mathbb{Z}_n$ at the SUSY vacuum, \Eref{vevs}.

\subsection{Framework of Inflationary Analysis}\label{sugra3}

To consolidate the validity of the inflationary proposal we have
to check the stability of the inflationary direction
\beq \th=s=\bar s=0,\label{inftr} \eeq
w.r.t the fluctuations of the various fields, which are expanded
in real and imaginary parts as follows
\beq \Phi= \frac{\phi}{\sqrt{2}}\,e^{i
\th/\mP}\>\>\>\mbox{and}\>\>\>S= \frac{s +i\bar
s}{\sqrt{2}}\cdot\label{cannor} \eeq
To this end we examine the validity of the extremum and minimum
conditions, i.e.,
\beqs\beq \left.{\partial
\Vhio\over\partial\what\chi^\al}\right|_{\mbox{\Eref{inftr}}}=0\>\>\>
\mbox{and}\>\>\>\what m^2_{
\chi^\al}>0\>\>\>\mbox{with}\>\>\>\chi^\al=\th,s,\bar
s.\label{Vcon} \eeq
Here $\what m^2_{\chi^\al}$ are the eigenvalues of the mass matrix
with elements
\beq \label{wM2}\what
M^2_{\al\bt}=\left.{\partial^2\Vhio\over\partial\what\chi^\al\partial\what\chi^\beta}\right|_{\mbox{\Eref{inftr}}}
\mbox{with}\>\>\>\chi^\al=\th,s,\bar s\eeq\eeqs
and hat denotes the EF canonically normalized fields. The kinetic
terms of the various scalars in \Eref{Saction1} can be brought
into the following form
\beqs\beq \label{K3} K_{\al\bbet}\dot z^\al \dot
z^{*\bbet}=\frac12\lf\dot{\se}^{2}+\dot{\what
\th}^{2}\rg+\frac12\lf\dot{\what s}^2 +\dot{\what{\overline
s}}^2\rg,\eeq
where the dot denotes derivation w.r.t the JF cosmic time and the
hatted fields are defined as follows
\beq  \label{cannor3b} {d\widehat \sg\over
d\sg}=J=\sqrt{K_{\Phi\Phi^*}},\>\>\> \what{\th}=
\mP\sqrt{K_{\Phi\Phi^*}}\,\th/\phi,\>\>\>\mbox{and}\>\>\>(\what
s,\what{\bar s})=\sqrt{K_{SS^*}} {(s,\bar s)}.\eeq\eeqs
Note, in passing, that the spinors $\psi_\Phi$ and $\psi_S$
associated with the superfields $S$ and $\Phi$ are normalized
similarly, i.e., $\what\psi_{S}=\sqrt{K_{SS^*}}\psi_{S}$ and
$\what\psi_{\Phi}=\sqrt{K_{\Phi\Phi^*}}\psi_{\Phi}$.

Upon diagonalization of $\what M^2_{\al\bt}$, \Eref{wM2}, we can
construct the scalar mass spectrum of the theory along the
direction in \Eref{inftr} -- see \Sref{fhi11} and \ref{fgi1}.
Besides the stability requirement in \Eref{Vcon}, from the derived
spectrum we can numerically verify that the various masses remain
greater than $\Hhi$ during the last $50$ e-foldings of inflation,
and so any inflationary perturbations of the fields other than the
inflaton are safely eliminated. Due to the large effective masses
that $\th,s$ and $\bar s$ in \Eref{wM2} acquire during inflation,
they enter a phase of oscillations about zero with reducing
amplitude. As a consequence, the $\sg$ dependence in their
normalization -- see \Eref{cannor3b} -- does not affect their
dynamics. Moreover, we can observe that the fermionic (4) and
bosonic (4) degrees of freedom are equal -- here we take into
account that $\what\phi$ is not perturbed. Employing the
well-known Coleman-Weinberg formula \cite{cw}, we find that the
one-loop corrected inflationary potential is
\beq\Vhi=\Vhio+{1\over64\pi^2}\lf \widehat m_{\th}^4\ln{\widehat
m_{\th}^2\over\Lambda^2} +2 \widehat m_{s}^4\ln{\widehat
m_{s}^2\over\Lambda^2}-4\widehat
m_{\psi_{\pm}}^4\ln{m_{\widehat\psi_{\pm}}^2\over\Lambda^2}\rg
,\label{Vhic}\eeq
where $\Lambda$ is a renormalization group  mass scale,  $\widehat
m_{\th}$ and $\widehat m_{s}=\widehat m_{\bar s}$ are defined in
\Eref{Vcon} and $\what m_{\psi_{\pm}}$ are the mass eigenvalues
which correspond to eigenstates
$\widehat\psi_{\pm}\simeq(\what\psi_{S}\pm\what\psi_{\Phi})/\sqrt{2}$.
As we numerically verify, the one-loop corrections have no impact
on our results, since the slope of the inflationary path is
generated at the classical level and the various masses are
proportional to the weak coupling $\ld$.

\section{The Inflationary Scenaria}\label{fhi}

In this section we outline the salient features and the
predictions of our inflationary scenaria in Secs.~\ref{fhi1} and
\ref{fgi} respectively, testing them against a number of criteria
introduced in Sec.~\ref{fhi2}.

\subsection{Inflationary Observables -- Constraints} \label{fhi2}

A successful inflationary scenario has to be compatible with a
number of observational requirements which are outlined in the
following.

\paragraph{\ftn\sf 3.1.1.} The number of e-folds, $\widehat N_\star$, that
the scale $k_\star=0.05/{\rm Mpc}$ suffers during IG inflation,
\begin{equation}
\label{Nhi}  \Ne_\star=\int_{\se_{\rm f}}^{\se_\star}\,
\frac{d\se}{m^2_{\rm P}}\: \frac{\Ve_{\rm IG}}{\Ve_{\rm IG,\se}}=
\int_{\sg_{\rm f}}^{\sg_\star}\, J^2\frac{\Ve_{\rm IG}}{\Ve_{\rm
IG,\sg}}{d\sg\over\mP^2},
\end{equation}
has to be at least enough to resolve the horizon and flatness
problems of standard big bang, i.e., \cite{plin}
\begin{equation}  \label{Ntot}
\widehat{N}_\star\simeq19.4+2\ln{\what V_{\rm
IG}(\sg_\star)^{1/4}\over{1~{\rm GeV}}}-{4\over 3}\ln{\what V_{\rm
IG}(\sg_{\rm f})^{1/4}\over{1~{\rm GeV}}}+ {1\over3}\ln {T_{\rm
rh}\over{1~{\rm
GeV}}}+{1\over2}\ln{\fr(\sg_\star)\over\fr(\sg_{\rm f})^{1/3}},
\end{equation}
where we assumed that IG inflation is followed in turn by a
decaying-inflaton, radiation and matter domination, $\Trh$ is the
reheat temperature after IG inflation, $\sg_\star~[\se_\star]$ is
the value of $\sg~[\se]$ when $k_\star$ crosses outside the
inflationary horizon, and $\sg_{\rm f}~[\se_{\rm f}]$ is the value
of $\sg~[\se]$ at the end of IG inflation, which can be found, in
the slow-roll approximation and for the considered in this paper
models, from the condition
\beqs\beq {\sf max}\{\widehat\epsilon(\sg_{\rm
f}),|\widehat\eta(\sg_{\rm f})|\}=1,\label{srcon}\eeq
where the slow-roll parameters can be calculated as follows:
\beq \label{sr}\widehat\epsilon= {\mP^2\over2}\left(\frac{\Ve_{\rm
IG,\se}}{\Ve_{\rm
IG}}\right)^2={\mP^2\over2J^2}\left(\frac{\Ve_{\rm
IG,\sg}}{\Ve_{\rm IG}}\right)^2
\>\>\>\mbox{and}\>\>\>\>\>\widehat\eta= m^2_{\rm P}~\frac{\Ve_{\rm
IG,\se\se}}{\Ve_{\rm IG}}={\mP^2\over J^2}\left(\frac{\Ve_{\rm
IG,\sg\sg}}{\Ve_{\rm IG}}-\frac{\Ve_{\rm IG,\sg}}{\Ve_{\rm
IG}}{J_{,\sg}\over J}\right)\cdot \eeq\eeqs

\paragraph{\ftn\sf 3.1.2.} The amplitude $A_{\rm s}$ of the power spectrum of the curvature perturbation
generated by $\sg$ at the pivot scale $k_\star$ must to be
consistent with data~\cite{plin}
\begin{equation}  \label{Prob}
\sqrt{A_{\rm s}}=\: \frac{1}{2\sqrt{3}\, \pi\mP^3} \;
\frac{\Ve_{\rm IG}(\sex)^{3/2}}{|\Ve_{\rm
IG,\se}(\sex)|}=\frac{|J(\sg_\star)|}{2\sqrt{3}\, \pi\mP^3} \;
\frac{\Ve_{\rm IG}(\sg_\star)^{3/2}}{|\Ve_{\rm
IG,\sg}(\sg_\star)|}\simeq4.685\cdot 10^{-5},
\end{equation}
where we assume that no other contributions to the observed
curvature perturbation exists.

\paragraph{\ftn\sf 3.1.3.}  The (scalar) spectral index, $n_{\rm s}$, its
running, $a_{\rm s}$, and the scalar-to-tensor ratio $r$ --
estimated through the relations:
\beq\label{ns} \ns=\: 1-6\widehat\epsilon_\star\ +\
2\widehat\eta_\star,\hspace*{0.2cm}
\as=\:2\left(4\widehat\eta_\star^2-(\ns-1)^2\right)/3-2\widehat\xi_\star\hspace*{0.2cm}
\mbox{and}\hspace*{0.2cm} r=16\widehat\epsilon_\star, \eeq
where $\widehat\xi=\mP^4 {\Ve_{\rm IG,\widehat\sg} \Ve_{\rm
IG,\widehat\sg\widehat\sg\widehat\sg}/\Ve_{}^2}= \mP^2\,\Ve_{\rm
IG,\sg}\,\widehat\eta_{,\sg}/\Ve_{\rm
IG}\,J^2+2\widehat\eta\widehat\epsilon$ and the variables with
subscript $\star$ are evaluated at $\sg=\sg_\star$ -- must be in
agreement with the fitting of the data \cite{plin} with
$\Lambda$CDM model, i.e.,
\begin{equation}\label{nswmap}
\mbox{\ftn\sf (a)}\>\>\> \ns=0.9603\pm0.0146,\>\>\>\mbox{\ftn\sf
(b)}\>\>\> -0.0314\leq \as\leq0.0046
\>\>\>\mbox{and}\>\>\>\mbox{\ftn\sf (c)}\>\>\>r<0.135,
\end{equation}
at 95$\%$ \emph{confidence level} (c.l.)

\paragraph{\ftn\sf 3.1.4.}  To avoid corrections from quantum
gravity and any destabilization of our inflationary scenario due
to higher order non-renormalizable terms -- see \Eref{fdef} --, we
impose two additional theoretical constraints on our models --
keeping in mind that $\Vhi(\sg_{\rm f})\leq\Vhi(\sg_\star)$:
\beq \label{subP}\mbox{\ftn\sf (a)}\hspace*{0.2cm}
\Vhi(\sg_\star)^{1/4}\leq\mP
\hspace*{0.2cm}\mbox{and}\hspace*{0.2cm}\mbox{\ftn\sf
(b)}\hspace*{0.2cm} \sg_\star\leq\mP.\eeq
As we show in \Sref{fhi3}, the UV cutoff of our model is $\mP$ and
so no concerns regarding the validity of the effective theory
arise.

\subsection{no-Scale SUGRA}\label{fhi1}

According to our analysis in \Sref{sugra2}, \FHI in the context of
no-scale SUGRA  can be achieved adopting a \Ka\ which depends at
least on two gauge singlet superfields -- the inflaton $\Phi$ and
an accompanying one $S$ -- and has the form
\beq \Khi=-3\mP^2\ln\lf
\fk(\Phi)+\fk^*(\Phi^{*})-{|S|^2\over3\mP^2}+{\kx}{|S|^4\over3\mP^4}\rg,\label{Kol}\eeq
as inferred by inserting \eqss{nsks}{Fkdef}{Omg} into \Eref{Omg1}.
Consequently, the \Km\ which corresponds to $K$ is $SU(2, 1)/SU(2)
\times U(1)_R\times \mathbb{Z}_n$ globally symmetric.  The
underlying symmetry of \Km\ allows us to avoid any mixing of
inflaton $\Phi$ with $S$ which fixes $\fsp=1$ -- see \Eref{frsp}.
We below extract the inflatonary potential in \Sref{fhi11} and
present our analytical and numerical results in \Sref{an1} and
\ref{num1} respectively.

\subsubsection{The Inflationary Potential}\label{fhi11}

Taking into account the form of $\fk,~\fr$ and $\fsp$ from
\eqs{fdef}{frsp}, \Eref{Vhig} reads
\beq \Vhio =\frac{\ld^2 \mP^4|1-2\fk|^2}{4\fr^2}=\frac{\ld^2 \mP^4
\ft^2}{4 \ck^4\xsg^{2n}},\label{Vhi00}\eeq
since $\fsp=1$ and $\fr=2\ck\xsg^n/2^{n/2}$ where we introduce the
dimensionless quantities
\beq \label{dimlss}
\xsg=\sg/\mP\>\>\>\mbox{and}\>\>\>\ft=2^{n/2-1} - \ck\xsg^n. \eeq
Obviously $\Vhio$ in \Eref{Vhi00} develops a plateau with almost
constant potential energy density corresponding to the Hubble
parameter
\beq \He_{\rm
IG}={\Vhio^{1/2}\over\sqrt{3}\mP}\simeq{\ld\mP\over2\sqrt{3}\ck}
\>\>\>\mbox{with}\>\>\>\Vhio\simeq\frac{\ld^2 \mP^4}{4
\ck^2}\,.\label{Vhio}\eeq Along the configuration of \Eref{inftr}
$K_{\al\bbet}$ defined in \Eref{K3} takes the form
\beq \lf K_{\al\bbet}\rg={1\over\fr}\diag\lf {3\mP^2|\Omega_{{\rm
H},\sg}|^2\over\fr},1\rg= \diag\lf
{3n^2\over2\xsg^2},{2^{n/2}\over2\ck\xsg^n}\rg, \label{VJe3}\eeq
where the explicit form of $\fk$ in \Eref{fdef} is taken into
account. Integrating the first equation in \Eref{cannor3b} we can
identify the EF field:
\beq \se=\se_{\rm
c}+\sqrt{\frac{3}{2}}n\mP\ln\frac{\sg}{\vev{\sg}}\>\>\>\mbox{with}
\>\>\>\vev{\phi}=\frac{\sqrt{2}\mP}{\sqrt[n]{2\ck}},\label{se1}\eeq
where we take into account \eqs{fdef}{vevs}. Also $\se_{\rm c}$ is
a constant of integration.

\renewcommand{\arraystretch}{1.4}
\begin{table}[!t]
\bec\begin{tabular}{|c|c|l|}\hline
{\sc Fields} &{\sc Eingestates} & \hspace*{3.cm}{\sc Masses Squared}\\
\hline \hline
$1$ real scalar &$\what{\th}$ & $\what m^2_{\th}=
\ld^2\mP^2(2^{n-2}-\ck^2\xsg^n\ft)/3\ck^4\xsg^{2n}\simeq4\He_{\rm IG}^2$\\
$2$ real scalars &$\what{s},~\what{\bar s}$ & $\what m^2_{
s}=\ld^2\mP^2 (2^{3n/2} + 4\ck\xsg^n (2^n - 2^{n/2}\ck\xsg^n+$\\
&&$+12\kx\ft^2))/3\cdot 2^{3+n/2}\ck^4\xsg^{2n}$\\\hline
$2$ Weyl spinors & $\what{\psi}_\pm={\what{\psi}_{\Phi}\pm
\what{\psi}_{S}\over\sqrt{2}}$& $\what m^2_{ \psi\pm}\simeq
2^{n-2}\ld^2\mP^2/3\ck^4\xsg^{2n}$\\ \hline
\end{tabular}\ec
\hfill \vchcaption[]{\sl\small The mass spectrum along the
trajectory of \Eref{inftr} during IG inflation.}\label{tab3}
\end{table}

Following the general analysis in \Sref{sugra3} we derive the mass
spectrum along the configuration of \Eref{inftr}. Our results are
arranged in \Tref{tab3}. We see there that $\kx\gtrsim1$ assists
us to achieve $\what m^2_{{s}}>0$ -- in accordance with
\cref{linde,eno7,zavalos}. Inserting the extracted masses in
\Eref{Vhic} we can proceed to the numerical analysis of \FHI in
the EF \cite{induced}, employing the standard slow-roll
approximation \cite{review} -- see \Sref{num1}. For the sake of
the presentation, however, we first -- see \Sref{an1} -- present
analytic results based on \Eref{Vhio}, which are quite close to
the numerical ones.

\subsubsection{Analytic Results}\label{an1}

The duration of the slow-roll \FHI is controlled by the slow-roll
parameters which, according to their definition in \Eref{sr}, are
calculated to be
\beq \label{sr1}\what \epsilon\simeq \frac{2^n}{3
\ft^2}\>\>\>\mbox{and}\>\>\>\what\eta\simeq\frac{2^{1+n/2}(2^{n/2}-\ck\xsg^n)}{3
\ft^2}\cdot\eeq
The termination of \FHI is triggered by the violation of the
$\what \epsilon$ criterion at $\sg=\sgf$ given by
\beqs\beq \what\epsilon\lf\sgf\rg=1\>\Rightarrow\>
\sgf=\sqrt{2}\mP\lf(\sqrt{3}+2)/2\sqrt{3}\ck\rg^{1/n},
\label{sgap}\eeq
since the violation of the $\what \eta$ criterion occurs at
$\sg=\tilde\sg_{f}$ such that
\beq \what\eta\lf\tilde\sg_{\rm f}\rg=1\>\Rightarrow\>
\tilde\sgf=\sqrt{2}\mP\lf{5\over6\ck}\rg^{1/n}=\lf(3+2\sqrt{3})/5\rg^{-1/n}\sgf<\sgf.
\label{sgap1}\eeq\eeqs
In the EF, $\se_{\rm f}$ remains independent of $\ck$ and $n$,
since substituting \Eref{sgap} into \Eref{se1} we obtain
\beq \se_{\rm f}-\se_{\rm c}\simeq\sqrt{{3/2}}\mP
\ln(1+2/\sqrt{3}).\eeq E.g., setting $\se_{\rm c}=0$, we obtain
$\se_{\rm f}=0.94\mP$.

Given that $\sgf\ll\sg_\star$, we can find a relation between
$\sg_\star$ and ${\Ne}_\star$ as follows
\beqs\beq \label{s*}
{\Ne}_\star\simeq{3\ck\over2^{1+n/2}\mP^n}\lf{\sg_\star^n-\sgf^n}\rg\>\Rightarrow\>
\sg_\star\simeq\mP\sqrt[n]{2^{1+n/2}\Ne_\star/3\ck}.\eeq
Obviously, \FHI consistent with \sEref{subP}{b} can be achieved if
\beq \label{fsub}
x_\star\leq1\>\>\>\Rightarrow\>\>\>\ck\geq2^{1+n/2}\Ne_\star/3
\>\>\>\mbox{with}\>\>\>x_\star=\sg_\star/\mP\,.\eeq\eeqs
Therefore, we need relatively large $\ck$'s which increase with
$n$. On the other hand, $\se_\star$ remains \trns, since plugging
\Eref{s*} into \Eref{se1} we find
\beq \se_\star\simeq\se_{\rm c}+\sqrt{3/2}\mP\ln(4\Ne_\star/3),
\label{se*}\eeq
which gives $\se_\star=5.3\mP$ for $\se_{\rm c}=0$. Despite this
fact, our construction remains stable under possible corrections
from non-renormalizable terms in $\fk$ since these are expressed
in terms of initial field $\Phi$ and can be harmless for
$|\Phi|\leq\mP$.

Upon substitution of \eqss{Vhio}{VJe3}{s*} into \Eref{Prob} we
find $A_{\rm s}$ as follows
\begin{equation}  A^{1/2}_{\rm s}=\frac{\ld
\fp(x_\star)^2}{2^{n/2+2}\sqrt{2}\pi \ck^2
x_\star^n}=\frac{\ld(3-4\Ne_\star)^2}{96\sqrt{2}\ck\pi\Ne_\star}\>\>\Rightarrow\>\>\ld\simeq6\pi\sqrt{2\As}\ck/\Ne_\star\>\>\Rightarrow\>\>
\ck\simeq41637\ld\,,\label{lan} \eeq
for $\Ne_\star\simeq52$. Therefore, enforcing \Eref{Prob} we
obtain a relation between $\ld$ and $\ck$ which turns out to be
independent of $n$. Replacing $\sg_\star$ by \Eref{s*} into
\Eref{ns} we estimate, finally, the inflationary observable
through the relations:
\beqs\baq \label{ns1} && n_{\rm s}=\frac{(1+4\Ne_\star)(4\Ne_\star-15)}{(3-4\Ne_\star)^2}\simeq1-{2/\what N_\star}-9/2\what N_\star^2=0.960, \>\>\> \\
&& \label{as} a_{\rm s} \simeq-2\what\xi_\star=\frac{128(3-\Ne_\star)}{(4\Ne_\star-3)^3}\simeq{-2/ \what N^2_\star}+3/2\what N^3_\star=-0.0007,\>\>\> \\
 && \label{rt} r=\frac{192}{(3-4\Ne_\star)^2}\simeq{12/\what
 N^2_\star}=0.0045
\eaq\eeqs
for $\Ne_\star\simeq52$. These outputs are fully consistent with
the observational data, \Eref{nswmap}.

\begin{figure}[!t]\vspace*{-.26in}
\begin{center}
\epsfig{file=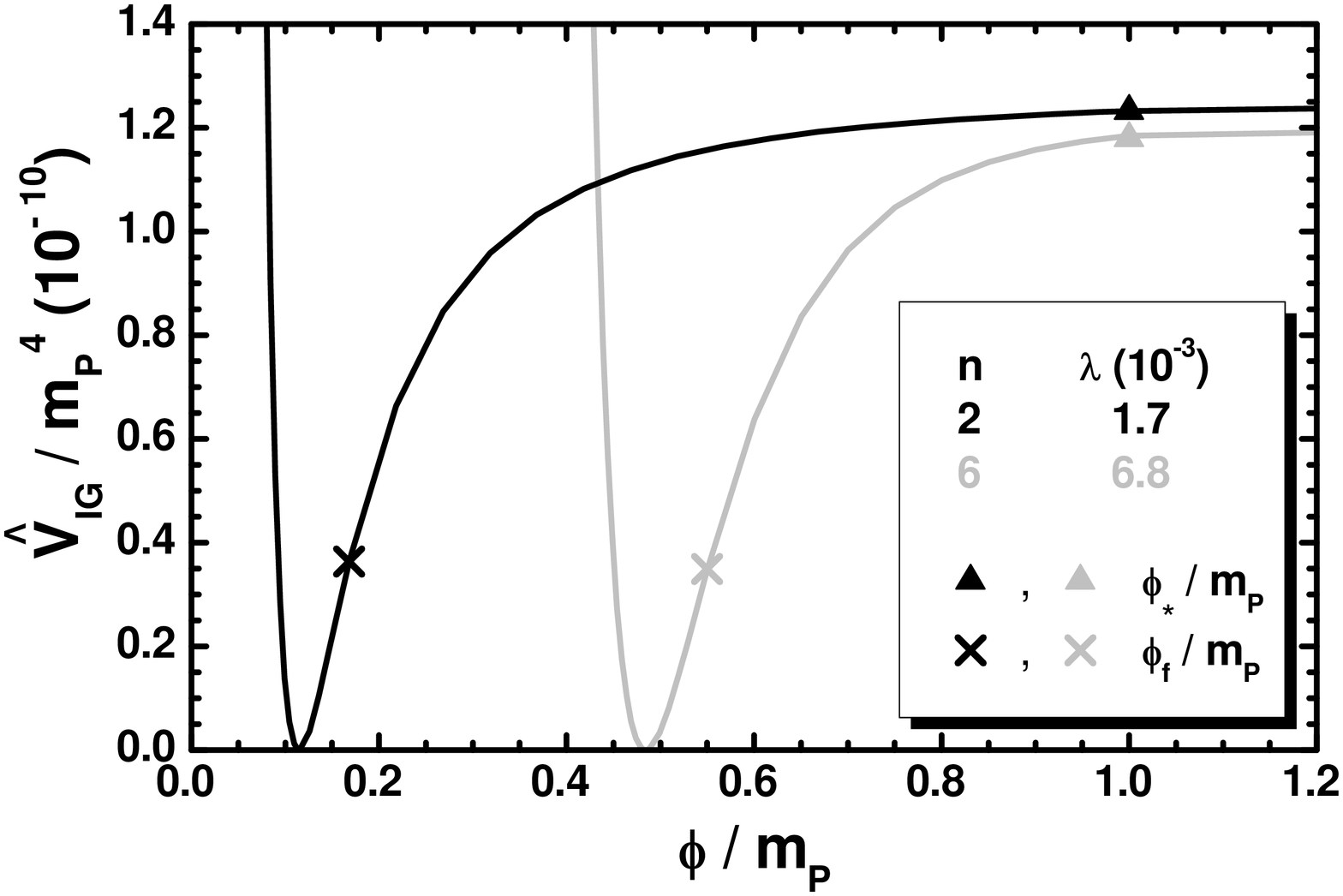,height=3.65in,angle=-90}\ec
\vspace*{-.15in}\hfill \vchcaption[]{\sl \small The inflationary
potential $\Vhi$ as a function of $\sg$ for $n=2$ and
$\ld=1.7\cdot10^{-3}$ (black line) or $n=6$ and
$\ld=6.8\cdot10^{-3}$ (light gray line). The values corresponding
to $\sgx$ and $\sgf$ are also depicted.}\label{fig3}
\end{figure}

\subsubsection{Numerical Results}\label{num1}

The inflationary scenario under consideration depends on the
parameters:
$$\ld,\>\ck,\>\kx\>\>\>\mbox{and}\>\>\>\Trh.$$
Our results are essentially independent of $\kx$'s, provided that
we choose them so as $\what m_{s}^2>0$ for every allowed $\ld$ and
$\ck$ -- see \Tref{tab3}. We therefore set $\kx=1$ throughout our
calculation. We also choose $\Lambda\simeq10^{13}~\GeV$ so as the
one-loop corrections in \Eref{Vhic} vanish at the SUSY vacuum,
\eqs{vevs}{ig}. Finally we choose $\Trh=10^9~\GeV$ as suggested by
reliable post-inflationary scenaria -- see \cref{pallis}. Upon
substitution of $\Vhi$ from \eqs{Vhic}{Vhio} in
\eqss{sr}{Nhi}{Prob} we extract the inflationary observables as
functions of $\ck$, $\ld$ and $\sg_\star$. The two latter
parameters can be determined by enforcing the fulfilment of
\Eref{Ntot} and (\ref{Prob}), for every chosen $\ck$. Our
numerical findings are quite close to the analytic ones listed in
\Sref{an1} for presentational purposes.

The variation of $\Vhi$ as a function of $\sg$ for two different
values of $n$ can be easily inferred  from \Fref{fig3}, where we
depict $\Vhi$ versus $\sg$ for $\sg_\star=\mP$ and $n=2$ (black
line) or $n=6$ (light gray line). The imposition  $\sg_\star=\mP$
corresponds to $\ld=0.0017$ and $\ck=76$ for $n=2$ and
$\ld=0.0068$ and $\ck=310$ for $n=6$. In accordance with our
findings in \eqs{se1}{fsub} we conclude that increasing $n$
{\sf\ftn (i)} larger $\ck$'s and therefore lower $\Vhio$'s are
required to obtain $\sg<\mP$; {\sf\ftn (ii)} larger $\sgf$ and
$\vev{\phi}$ are obtained. Combining \eqs{sgap}{lan} with
\Eref{Vhio} we can convince ourselves that $\Vhio(\sgf)$ is
independent of $\ck$ and to a considerable degree of $n$.

\begin{figure}[!t]\vspace*{-.12in}
\hspace*{-.19in}
\begin{minipage}{8in}
\epsfig{file=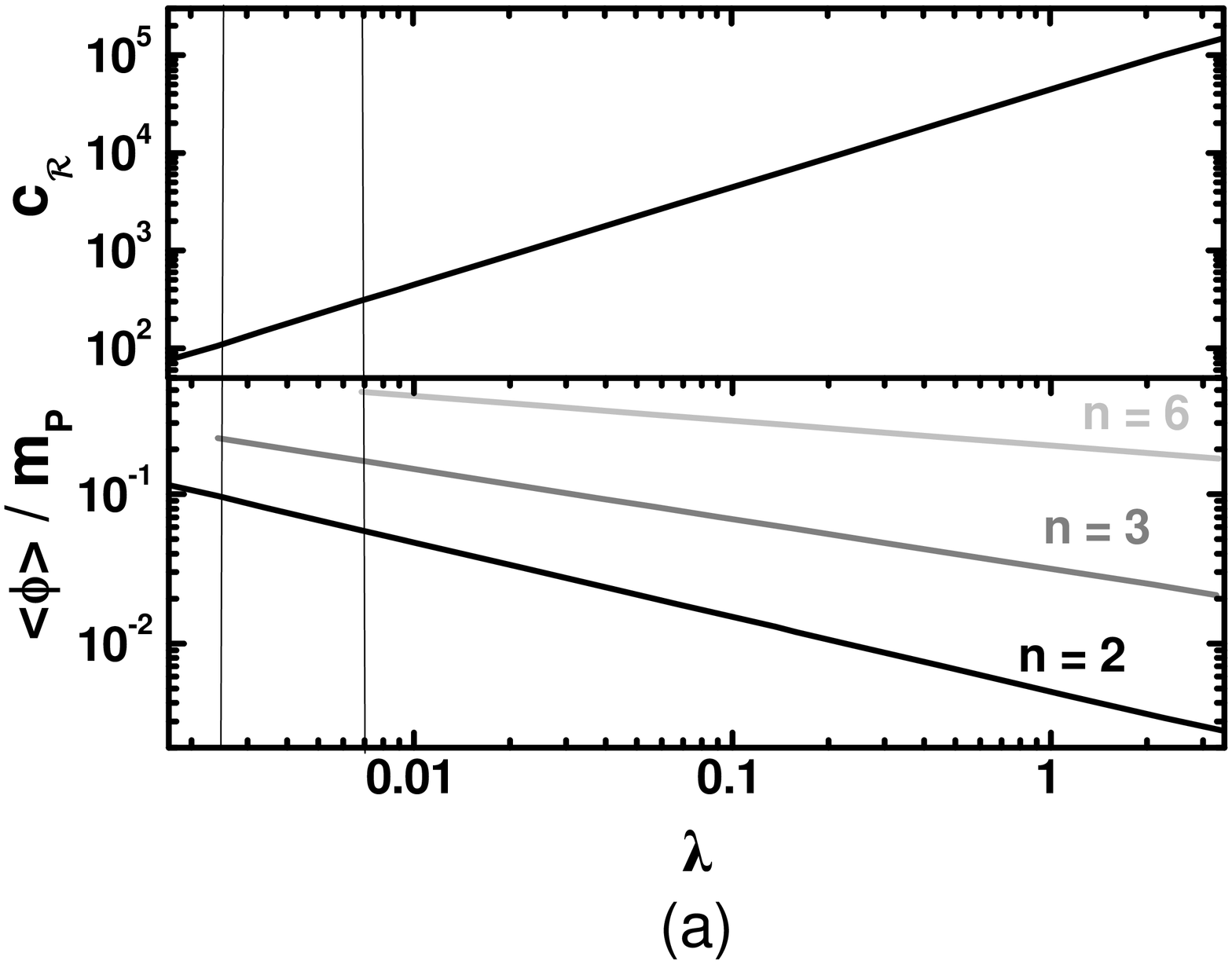,height=3.6in,angle=-90}
\hspace*{-1.2cm}
\epsfig{file=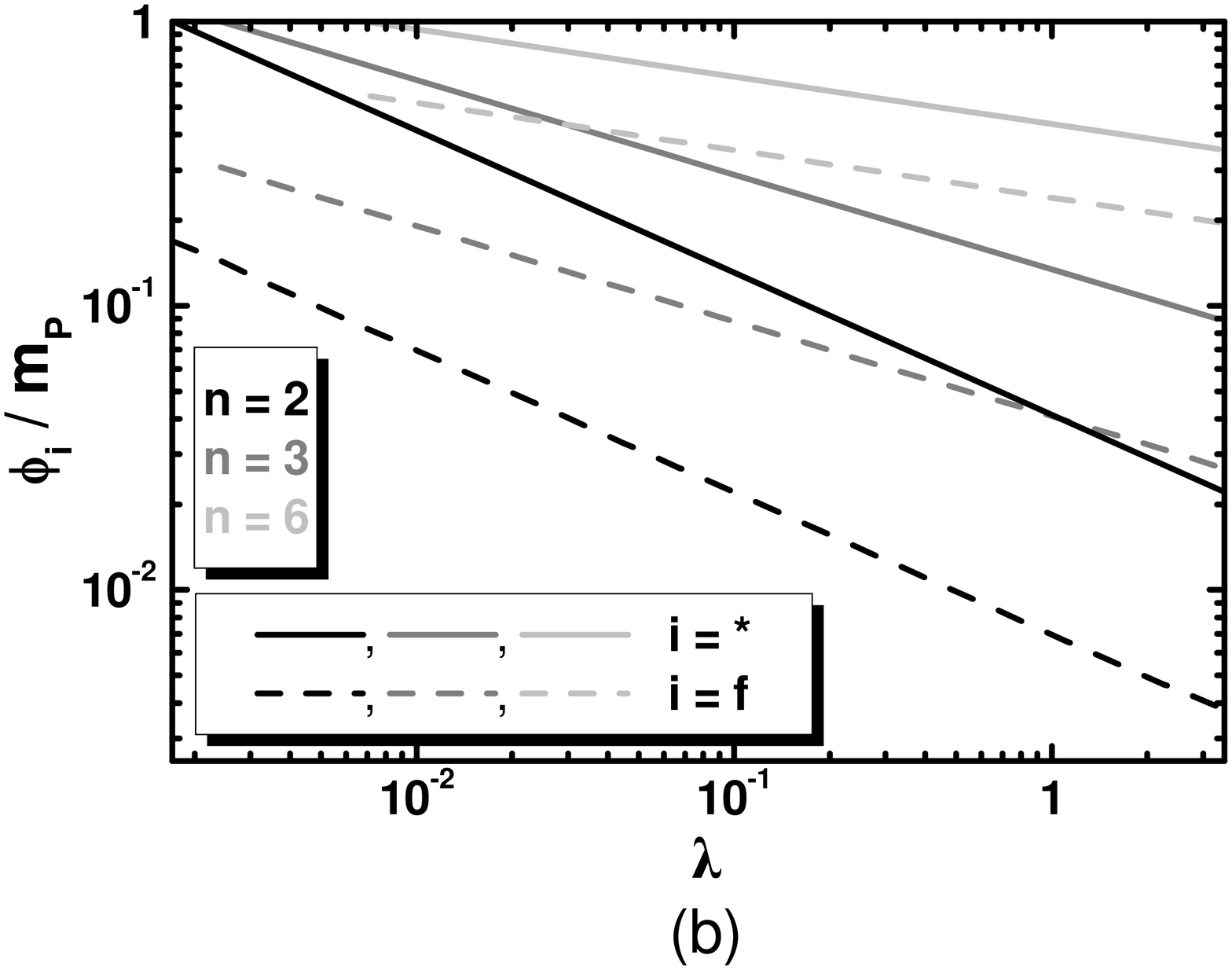,height=3.6in,angle=-90} \hfill
\end{minipage}
\hfill \vchcaption[]{\sl\small   The allowed by Eqs.~(\ref{Ntot}),
(\ref{Prob}) and (\ref{subP}) values of $\ck$ and the resulting
$\vev{\phi}$ [$\sg_\star$ (solid line) and $\sg_{\rm f}$ (dashed
line)] versus $\ld$ (a) [(b)]. We use black, gray and light gray
lines for $n=2,3$ and $6$ respectively, $\kx=1$ and
$\Trh=10^{9}~\GeV$. \Eref{subP} is fulfilled to the right of the
thin line.}\label{fig1}
\end{figure}


By varying $\ld$ we can delineate the region of the parameters
allowed by a simultaneous imposition of \eqss{Prob}{Ntot}{subP}.
Our results are displayed in \Fref{fig1}, where we draw as
functions of $\ld$ the allowed values of $\ck$ and $\vev{\phi}$ --
see \sFig{fig1}{a} -- $\sg_\star$ (solid line) and $\sg_{\rm f}$
(dashed line) -- see \sFig{fig1}{b}. We use black, gray and light
gray lines for $n=2,3$ and $6$ respectively. As anticipated in
\Eref{lan} the relation between $\ck$ and $\ld$ is independent of
$n$; the various lines, thus, coincide. However, \Eref{subP} is
fulfilled to the right of the thin line. Indeed, the lower bound
of the depicted lines comes from the saturation of \Eref{fsub}
whereas the upper bound originates from the perturbative bound on
$\ld$, $\ld\leq\sqrt{4\pi}\simeq3.54$. Moreover, the variation of
$\sg_{\rm f}$ and $\sg_\star$ as a function of $\ld$ -- drawn in
\sFref{fig1}{b} -- is consistent with \eqs{sgap}{s*}.

The overall allowed parameter space of the model for $n=2,3$ and
$6$ is correspondingly
\beqs\beq\label{res1} 76,105,310\lesssim
\ck\lesssim1.5\cdot10^5\>\>\>\mbox{and}\>\>\>(1.7,2.4,6.8)\cdot10^{-3}
\lesssim \ld\lesssim3.54\>\>\>\mbox{for}\>\>\>\Ne_\star\simeq52
\eeq with $\vev{\phi}$ being confined in the ranges
$(0.0026-0.1)$, $(0.021-0.24)$ and $(0.17-0.48)$. Moreover, the
masses of the various scalars in \Tref{tab3} remain well above
$\Hhi$ both during and after \FHI for the selected $\kx$. E.g.,
for $n=3$ and $\ck=495$ (corresponding to $\ld=0.01$) we obtain
\beq\label{res3} \lf \what m_{\th}^2(\sgx), \what
m_{s}^2(\sgx)\rg/\Hhi^2(\sgx)\simeq(4,905)\>\>\>\mbox{and}\>\>\>\lf
\what m_{\th}^2(\sgf), \what
m_{s}^2(\sgf)\rg/\Hhi^2(\sgf)\simeq(10.5,26.8). \eeq \eeqs %
Letting $\ld$ or $\ck$ vary within its allowed region in
\Eref{res1}, independently of $n$, we obtain
\beq\label{res} 0.961\lesssim \ns\lesssim0.963,\>\>\>-7\lesssim
{\as/10^{-4}}\lesssim-6.4\>\>\>\mbox{and}\>\>\>4.2\gtrsim
{r/10^{-3}}\gtrsim3.6,\eeq
which lie close to the analytic results in \eqss{ns1}{as}{rt} and
within the allowed ranges of \Eref{nswmap}, with $\ns$ being
impressively spot on its central observationally favored value --
see \sEref{nswmap}{a}. Therefore, the inclusion of the variant
exponent $n\geq2$, compared to the initial model of \cref{pallis},
does not affect the successful predictions on the inflationary
observables.

\subsection{Beyond no-Scale SUGRA}\label{fgi}

If we lift the assumption of no-scale SUGRA in \Eref{nsks},
$\Omega$ takes its more general form, obtained by inserting
\eqs{Fkdef}{fdef} into \Eref{Omg}; the resulting through
\Eref{Omg1} \Ka\ is
\beq  \Khi=-3\mP^2\ln\lf
\fk(\Phi)+\fk^*(\Phi^{*})-{|S|^2\over3\mP^2}-{|\Phi|^2\over3\mP^2}+{\kx}{|S|^4\over3\mP^4}+
2\kpp{|\Phi|^4\over3\mP^4}+2\ksp
{|S|^2|\Phi|^2\over3\mP^4}\rg,\label{Kolg}\eeq
where the factors of $2$ are added just for convenience. The
description of the inflationary potential, our analytical and
numerical results are exhibited below in Secs.~\ref{fgi1},
\ref{gnum1} and \ref{gnum2} correspondingly.

\subsubsection{The Inflationary Potential}\label{fgi1}

The tree-level scalar potential in this case has its general form
in \Eref{Vhig} where $\fr$ and $\fsp$ are calculated by employing
their definitions in \Eref{frsp} as follows
\beq
\fr=2\ck{\xsg^n\over2^{n/2}}+{\xsg^2\over6}+{\kpp\over12}\xsg^4\>\>\>\mbox{and}\>\>\>\fsp=1-\ksp\xsg^2.\label{fs3}
\eeq
Taking into account the form of $\fr$ above, $\Vhio$ can be cast
as follows
\beqs\beq
\label{3Vhioo}\Vhio=\frac{\ld^2\mP^4\ft^2}{4\ck^2\xsg^4(\ck
\xsg^{n-2}-2^{n/2-2} f_{\phi\phi}/3)^2\fsp},\eeq
%
where $f_{\phi\phi}=1-\kpp\xsg^2$ while $\xsg$ and $\ft$ are
defined in \Eref{dimlss}. Similarly to \Sref{fhi1}, $\Vhio$ in
\Eref{3Vhioo} develops a plateau with almost constant potential
energy density corresponding to the Hubble parameter
\beq \He_{\rm
IG}={\Vhio^{1/2}\over\sqrt{3}\mP}\simeq{\ld\mP\over2\sqrt{3\fsp}\ck}
\>\>\>\mbox{with}\>\>\>\Vhio\simeq\frac{\ld^2 \mP^4}{4
\fsp\ck^2}\,\cdot\label{3Vhio}\eeq\eeqs Moreover, the EF
canonically normalized inflaton, $\se$, is found via
\Eref{cannor3b} with $J^2$ given by
\beq \label{Jg}
J^2={3\over2}\frac{n^2\ck^2\xsg^{2n}+2^{4+n/2}\ck\xsg^{2+n}(1-n+2\kpp(n-2)\xsg^2)}
{(\ck\xsg^{1+n}-2^{n/2-2}\xsg^3f_{\phi\phi}/3)^2}\simeq{3n^2\over2\xsg^2}+{2^{n/2}(1-n)\over2\ck\xsg^n}\cdot\eeq
Consequently, $J$ turns out to be close to that obtained in
\Sref{fhi11}.

Following the standard procedure of \Sref{sugra3} we construct the
mass spectrum of the theory along the path of \Eref{inftr}. The
precise expressions of the relevant masses squared, taken into
account in our numerical computation, are rather lengthy due to
the numerous contributions to $\Vhio$, \Eref{3Vhioo}. Our
findings, though, can be considerably simplified, if we perform an
expansion for small $\xsg$'s -- retaining $\ft$ intact --,
consistently with our restriction, \Eref{subP}. If we keep the
lowest order terms, the masses squared for the scalars reduce to
those displayed in \Tref{tab3}, whereas the mass squared of the
chiral fermions shown in \Tref{tab3} has to be multiplied by the
factor
\beq 1+{\ksp\ck\xsg^{2+n}}/{2^{n/2-1}n}.\label{fma}\eeq
As in the case of \Sref{fhi1}, employing the mass spectrum along
the direction of \Eref{inftr}, we can calculate $\Vhi$ in
\Eref{Vhic} to further analyze the model.

\subsubsection{Analytic Results}\label{gnum1}

Upon substitution of \eqs{3Vhio}{Jg} into \Eref{sr}, we can
extract the slow-roll parameters which determine the strength of
the inflationary stage. Performing expansions about $\xsg\simeq0$,
we can achieve approximate expressions which assist us to
interpret the numerical results presented below. Namely, we find
\beq \nonumber\what\epsilon= \frac{(2^{n/2} n + 2 \ksp \ck \xsg^{2
+ n})^2}{3 n^2 \ft^2}\>\>\>\mbox{and}\>\>\>\what\eta={1\over 3 n^2
\ft^2}\times\eeq \beq\label{gsr1} { \lf 2^{n} n^2 + 4 \ksp \ck^2
\xsg^{2(1+n)} + 2^{n/2} \ck \xsg^n \lf\lf(n-2)^2/6 + 4 \ksp
(n-1)\rg \xsg^2-n^2\rg\rg}.\eeq
%
As it may be numerically verified, $\sgx\equiv x_\star\mP$ and
$\sgf$ do not decline a lot from their values in \eqs{s*}{sgap},
which can be served for our estimations below. In particular,
replacing $\Vhio$ from \Eref{3Vhio} in \Eref{Prob} we obtain
\begin{equation} \As^{1/2}=\frac{n\ld\ft^2(x_\star)}{4\sqrt{2}\pi\ck^2 x^n_\star(2^{n/2} n + 2 \ksp
\ck x^{2 + n}_\star)}
\>\>\Rightarrow\>\>\ld\simeq2\pi\sqrt{2\As}\ck\lf{3\over\Ne_\star}+{8\ksp\over
n}\lf{2\Ne_\star\over3\ck}\rg^{2/n}\rg\cdot \label{lang} \eeq
Comparing this expression with the one obtained in the case of
no-scale SUGRA, \Eref{lan}, we remark that $\ld$ acquires a mild
dependence on both $\ksp$ and $n$. Inserting \Eref{s*} into
\eqs{gsr1}{ns} we can similarly provide an expression for $n_{\rm
s}$. This is
\beq \label{gns}  n_{\rm s}\simeq1-{2\over\what N_\star}\ +\
\lf\frac{4}{9}\rg^{1/n}\lf\frac{\Ne_\star}{\ck}\rg^{2/n} \frac{128
\ksp + 27 n^2/\Ne_\star^3}{12 n^2}\cdot \eeq
Therefore, a clear dependence of $\ns$ on $n$ and $\ksp$ arises,
with the second one being much more efficient. On the other hand,
$\as$ and $r$ remain pretty close to those obtained in \Sref{an1}
-- see \eqs{as}{rt}. In particular, the dependence of $r$ on $n$
and $\ksp$ can be encoded as follows
\beq r\simeq {12\over\Ns^2}\ +\
 32\,  {2^{2/n + 1}\ksp \over
 3^{2/n}n \Ns^{1-2/n}\ck^{2/n}}\ +\ 64\, {2^{4/n + 2}\ksp^2 \Ns^{4/n}\over 3^{(4 + n)/n}n^2\ck^{4/n} } \cdot\label{ra} \eeq
It is clear from the results above that $\ksp\neq0$ has minor
impact on $r$ since its presence is accompanied by large
denominators where $\ck\gg1$ is involved.

\begin{figure}[!t]\vspace*{-.12in}
\hspace*{-.19in}
\begin{minipage}{8in}
\epsfig{file=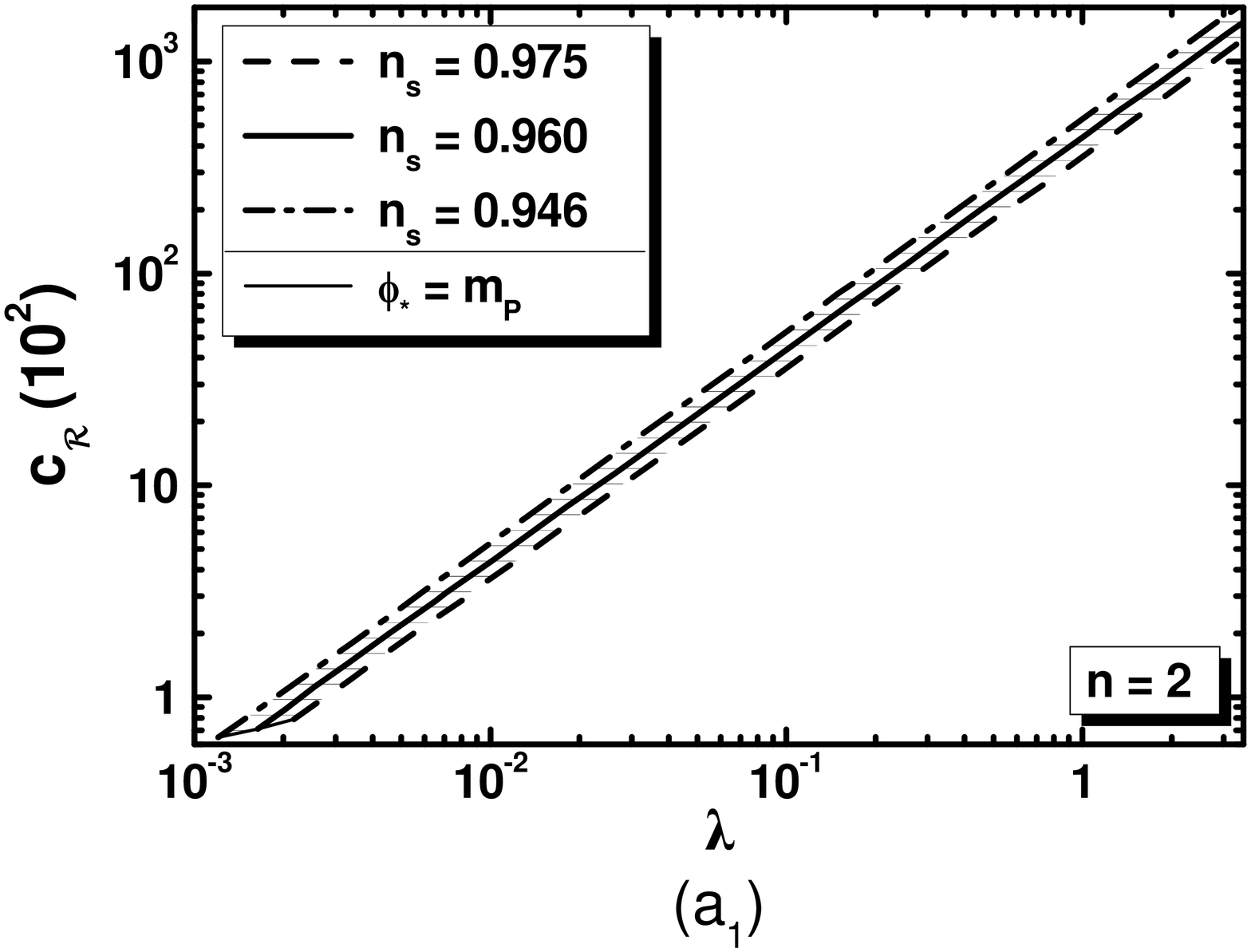,height=3.6in,angle=-90}
\hspace*{-1.2cm}
\epsfig{file=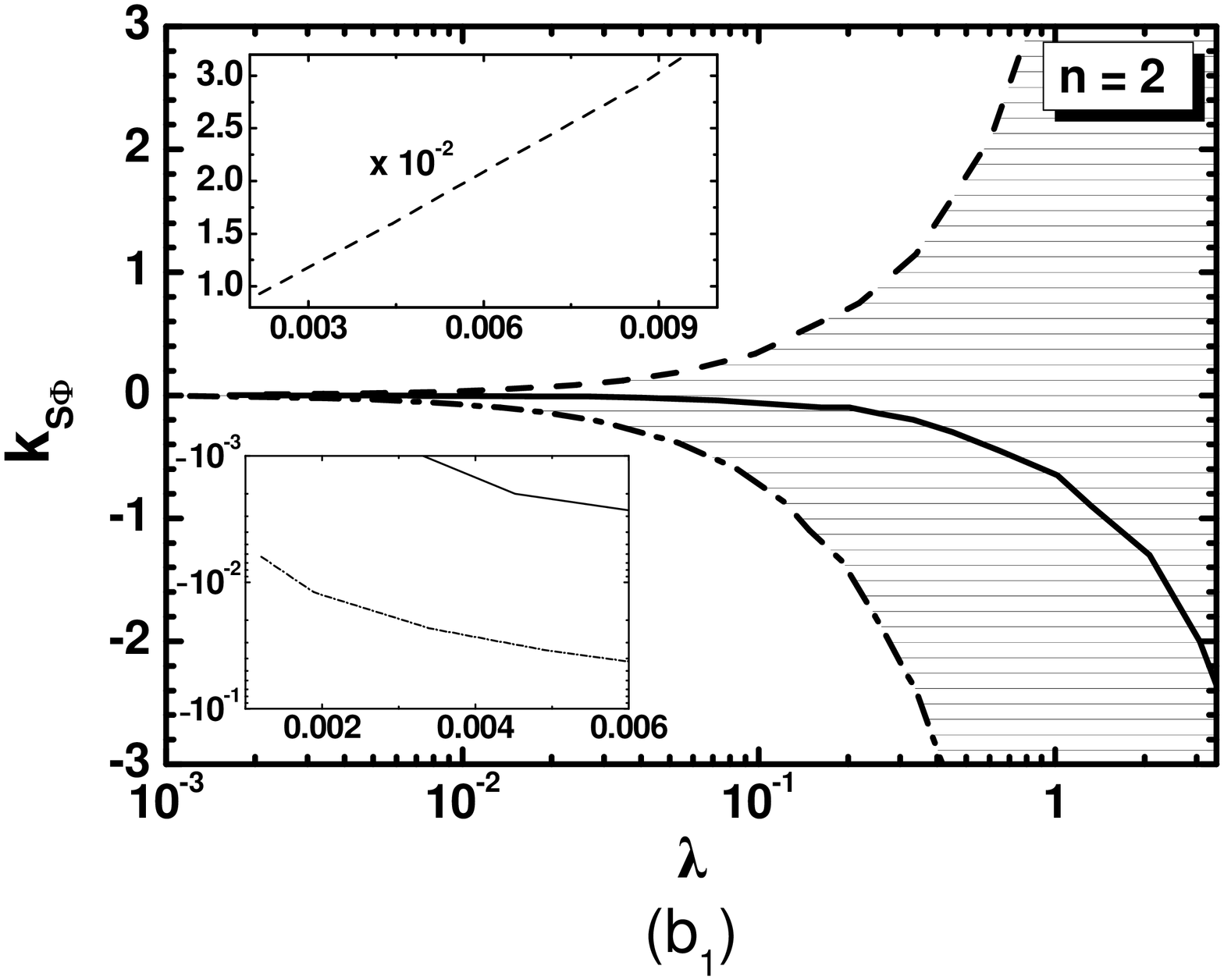,height=3.6in,angle=-90} \hfill
\end{minipage}
\hfill \hspace*{-.19in}
\begin{minipage}{8in}
\epsfig{file=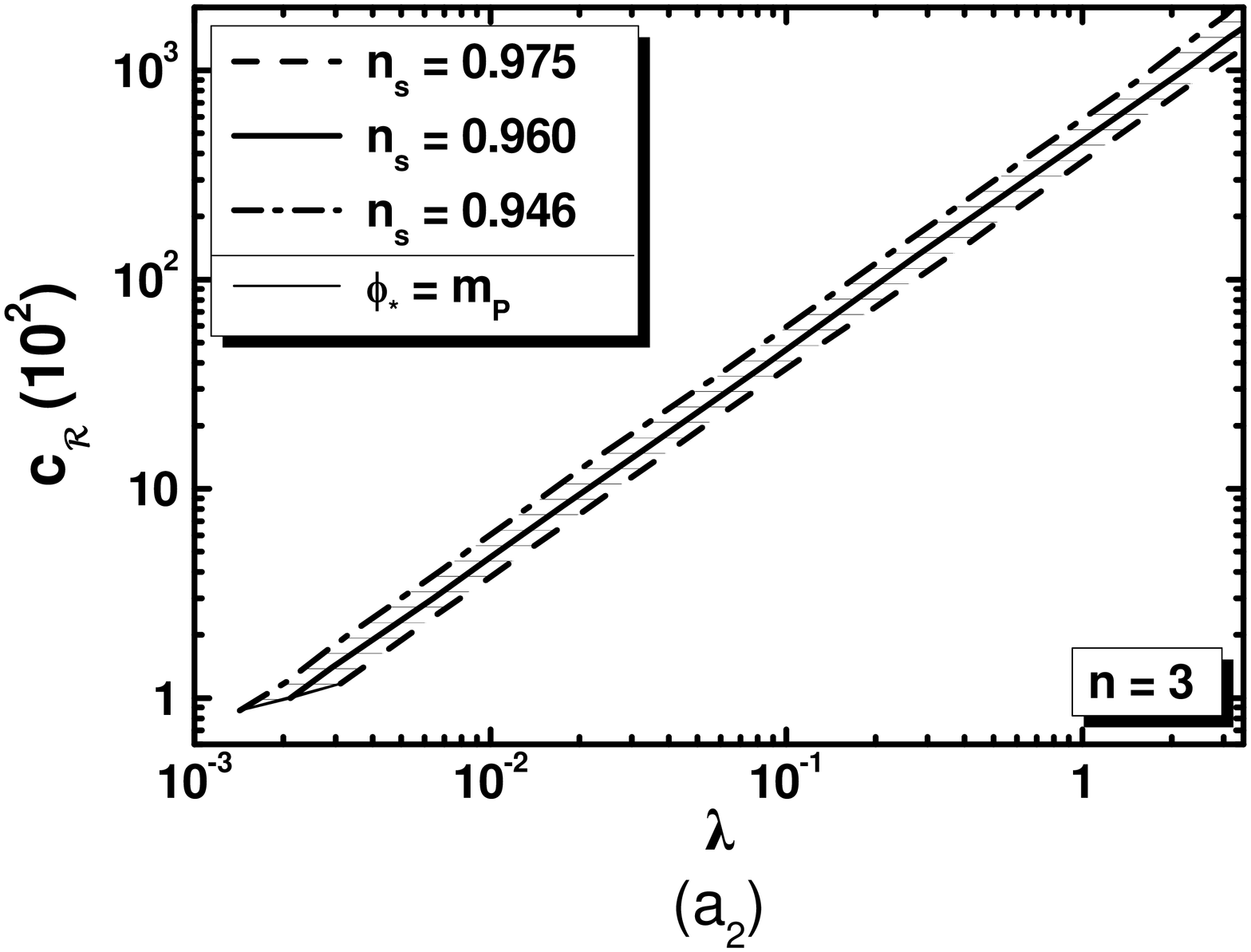,height=3.6in,angle=-90}
\hspace*{-1.2cm}
\epsfig{file=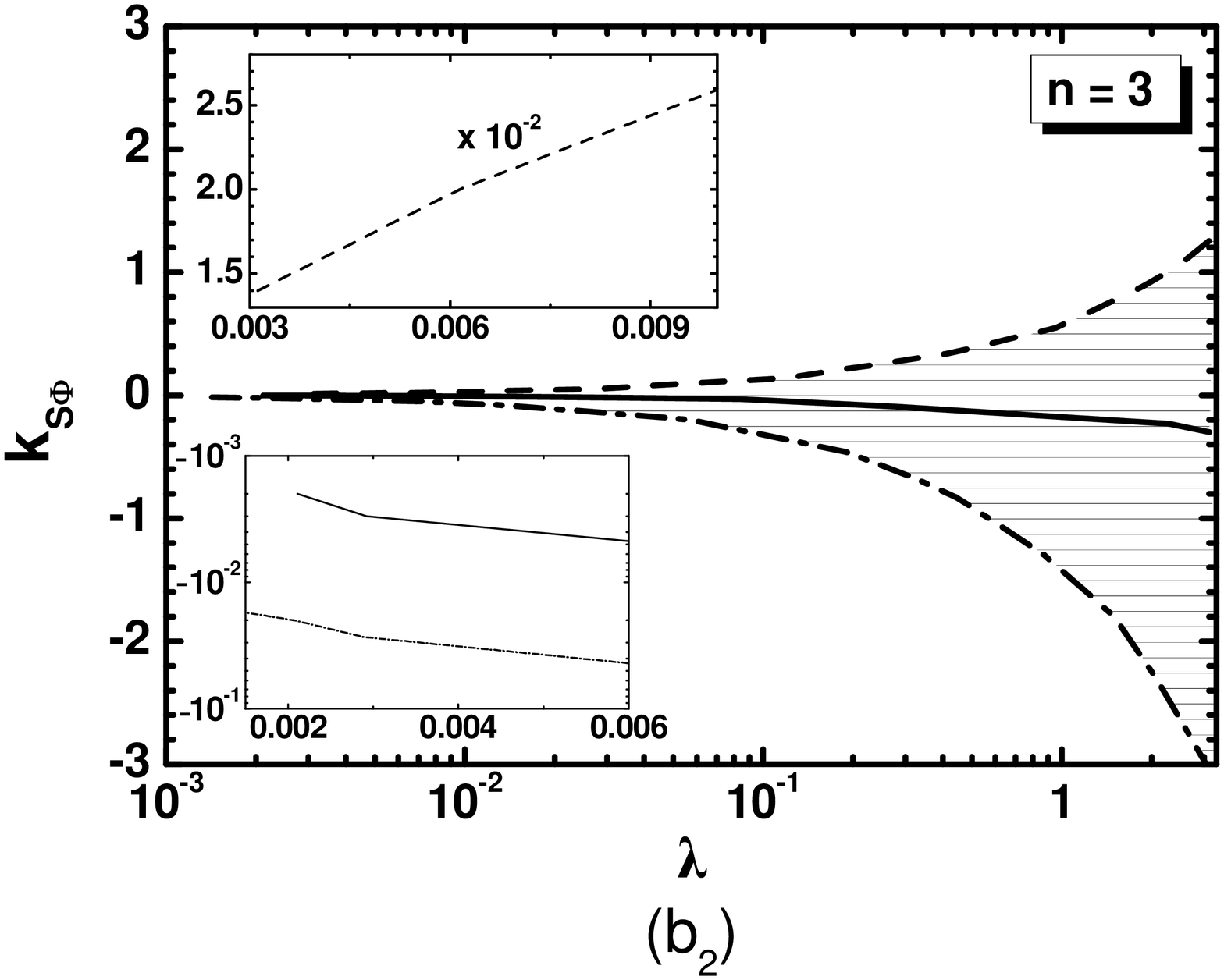,height=3.6in,angle=-90} \hfill
\end{minipage}
\hfill \vchcaption[]{\sl\small The (hatched) regions allowed by
Eqs.~(\ref{Ntot}), (\ref{Prob}), (\ref{nswmap}) and (\ref{subP})
in the $\ld-\ck$ plane (${\sf\ftn a}_1$, ${\sf\ftn a}_2$) and
$\ld-\ksp$ plane (${\sf\ftn b}_1$, ${\sf\ftn b}_2$) for $\ks=1$,
$\kpp=0.5$ and $n=2$ (${\sf\ftn a}_1$, ${\sf\ftn b}_1$) or $n=3$
(${\sf\ftn a}_2$, ${\sf\ftn b}_2$). The conventions adopted for
the various lines are also shown.}\label{fig2g}
\end{figure}


\subsubsection{Numerical Results}\label{gnum2}

This inflationary scenario depends on the following parameters:
$$\ld,\>\ck,\>\kx,\>\ksp,\>\kpp\>\>\>\mbox{and}\>\>\>\Trh.$$
As in the case of \Sref{num1} our results are independent of
$\ks$, provided that $\what  m_{s}^2>0$ -- see in \Tref{tab3}. The
same is also valid for $\kpp$ since the contribution from the
second term in $\fr$, \Eref{fs3}, is overshadowed by the strong
enough first term including $\ck\gg1$. We therefore set $\kx=1$
and $\kpp=0.5$. We also choose $\Trh=10^9~\GeV$. Besides these
values, in our numerical code, we use as input parameters
$\ck,~\ksp$ and $\sg_\star$. For every chosen $\ck\geq1$, we
restrict $\ld$ and $\sg_\star$ so that the conditions
\eqss{Nhi}{Prob}{subP} are satisfied. By adjusting $\ksp$ we can
achieve $\ns$'s in the range of Eq.~(\ref{nswmap}). Our results
are displayed in \Fref{fig2g}-({\sf\ftn a}$_1$) and ({\sf\ftn
a}$_2$) [\Fref{fig2g}-({\sf\ftn b}$_1$) and ({\sf\ftn b}$_2$)],
where we delineate the hatched regions allowed by
\eqsss{Nhi}{Prob}{nswmap}{subP} in the $\ld-\ck$ [$\ld-\ksp$]
plane. We take $n=2$ in \Fref{fig2g}-({\sf\ftn a}$_1$) and
({\sf\ftn b}$_1$) and $n=3$ in \Fref{fig2g}-({\sf\ftn a}$_2$) and
({\sf\ftn b}$_2$). The conventions adopted for the various lines
are also shown. In particular, the dashed [dot-dashed] lines
correspond to $n_{\rm s}=0.975$ [$n_{\rm s}=0.946$], whereas the
solid (thick) lines are obtained by fixing $n_{\rm s}=0.96$ -- see
Eq.~(\ref{nswmap}). Along the thin line, which provides the lower
bound for the regions presented in \Fref{fig2g}, the constraint of
\sEref{subP}{b} is saturated. At the other end, the perturbative
bound on $\ld$ bounds the various regions.

From \Fref{fig2g}-({\sf\ftn a}$_1$) and ({\sf\ftn a}$_2$) we see
that $\ck$ remains almost proportional to $\ld$ and for constant
$\ld$, $\ck$ increases as $\ns$ decreases. From
\Fref{fig2g}-({\sf\ftn b}$_1$) we remark that $\ksp$ is confined
close to zero for $n_{\rm s}=0.96$ and $\ld<0.16$ or
$\sg_\star>0.1\mP$ -- see \Eref{s*}. Therefore, a degree of tuning
(of the order of $10^{-2}$) is needed in order to reproduce the
experimental data of \sEref{nswmap}{a}. On the other hand, for
$\ld>0.16$ (or $\sg_\star<0.1\mP$), $\ksp$ takes quite natural (of
order one) negative values -- consistently with \Eref{gns}. This
feature, however, does not insist for $n=3$ -- see
\Fref{fig2g}-({\sf\ftn b}$_2$) --, where the allowed (hatched)
region is considerably shrunk and so, $\ksp$ remains constantly
below unity for any $\lambda$. As we explicitly verified, for
$n=6$ the results turn out to be even more concentrated about
$\ksp\simeq0$. Therefore, we can conclude that this embedding of
IG inflation in SUGRA favors low $n$ values.

More explicitly, for $n_{\rm s}=0.96$ and $\Ne_\star\simeq52$ we
find:
\beqs\bea\label{resg} && 71\lesssim
\ck\lesssim1.5\cdot10^5\>\>\>\mbox{with}\>\>\>1.6\cdot10^{-3}\lesssim
\ld\lesssim3.5\>\>\>\mbox{and}\>\>\> 0\lesssim -\ksp\lesssim2.4\,
\>\>\>(n=2);\>\>\>\>\>\>\\ && \label{resg2} 100\lesssim
\ck\lesssim1.4\cdot10^5\>\>\>\mbox{with}\>\>\>2.1\cdot10^{-3}\lesssim
\ld\lesssim3.5\>\>\>\mbox{and}\>\>\> 0.002\lesssim
-\ksp\lesssim0.3\,\>\>\>(n=3);\>\>\>\>\>\>\>\>\>\>\>\>\>\\ &&
\label{resg3}270\lesssim
\ck\lesssim1.65\cdot10^5\>\>\>\mbox{with}\>\>\>5.6\cdot10^{-3}\lesssim
\ld\lesssim3.5\>\>\>\mbox{and}\>\>\> 0.01\lesssim
-\ksp\lesssim0.1\,\>\>\>(n=6).\>\>\>\>\>\>\>\>\>\>\>\>\> \eea\eeqs
Note that the lower bounds on $\ck$ and $\ld$ are quite close to
those obtained in \Eref{res1}. In both cases  $6.8\lesssim
{|\as|/10^{-4}}\lesssim8.2$ and $r\simeq3.8\cdot 10^{-3}$ which
lie within the allowed ranges of \Eref{nswmap}. Needless to say
that, as in \Sref{num1}, we here also obtain $\what
m^2_{\chi^\al}/\Hhi^2\gg1$ with $\what m^2_{\chi^\al}$ being
defined in \Eref{Vcon}.

\section{The Effective Cut-off Scale}\label{fhi3}

An outstanding trademark of IG inflation is that it is
unitarity-safe, despite the fact that its implementation with
\sub\ $\phi$'s -- see \Eref{fsub} -- requires relatively large
$\ck$'s. To show this we below extract the UV cut-off scale,
$\Qef$, of the effective theory first in the JF -- \Sref{fhi3b} --
and then in the EF -- see \Sref{fhi3a}. Although the expansions
about $\vev{\phi}$ presented below  are not valid \cite{cutof}
during IG inflation, we consider the extracted this way $\Qef$ as
the overall cut-off scale of the theory, since reheating is an
unavoidable stage of the inflationary dynamics \cite{riotto}.

\subsection{Jordan Frame Computation}\label{fhi3b}

The possible problematic process in the JF, which causes
\cite{cutoff} concerns about the unitarity-violation, is the
$\dph-\dph$ scattering process via $s$-channel graviton,
$h^{\mu\nu}$, exchange -- $\dphi$ represents an excitation of
$\phi$ about $\vev{\phi}$, see below. The relevant vertex is
$\ck\dph^2\Box h/\mP$ -- with $h=h^\mu_{\mu}$ -- can be derived
from the first term in the right-hand side of \Eref{Sfinal}
expanding the JF metric $g_{\mu\nu}$ about the flat spacetime
metric $\eta_{\mu\nu}$ and the inflaton $\phi$ abound its v.e.v as
follows:
\beq
g_{\mu\nu}\simeq\eta_{\mu\nu}+h_{\mu\nu}/\mP\>\>\>\mbox{and}\>\>\>\phi=\vev{\phi}
+\dph.\eeq
Retaining only the terms with two derivatives of the excitations,
the part of the lagrangian corresponding to the two first terms in
the right-hand side of \Eref{Sfinal} takes the form
\beqs\baq \nonumber \delta{\cal L}&=&-{\vev{\fk}\over4}{F}_{\rm
EH}\lf h^{\mu\nu}\rg +\frac12\vev{F_{\rm K}}\partial_\mu
\dph\partial^\mu\dph+\lf\mP \vev{\Omega_{\rm H,\phi}}+\delta_{\cal
R}\ck^{2/n}{\dph\over\mP}\rg F_{\cal R}\dph\ +\cdots\\
&=&-{1\over8}F_{\rm EH}\lf \bar h^{\mu\nu}\rg+ \frac12\partial_\mu
\overline\dph\partial^\mu\overline\dph+\delta_{\cal
R}{\ck^{2/n}\over\sqrt{2}\mP}
\frac{\sqrt{\vev{\fk}}}{\vev{\bar\Omega_{\rm
H}}}\,\overline\dph^2\,\Box \bar h\, +\ \cdots,\label{L2}\eaq
where $\delta_{\cal R}=1/2$ [$\delta_{\cal R}=2^{2/n}n(n-1)/8$]
for $n=2$ [$n>2$] and the functions $F_{\rm EH}$, $F_{\cal R}$ and
$F_{\rm K}$ read
\bea && {F}_{\rm EH}\lf h^{\mu\nu}\rg= h^{\mu\nu} \Box
h_{\mu\nu}-h\Box h+2\partial_\rho h^{\mu\rho}\partial^\nu
h_{\mu\nu}-2\partial_\nu h^{\mu\nu}\partial_\mu h,\label{Leh}\\
&& F_{\cal R}\lf h^{\mu\nu}\rg=\Box h-
\partial_\mu \partial_\nu h^{\mu\nu}\label{Fr}\eea and \beq F_{\rm
K}=\left\{\bem
%
0,\hfill&\mbox{for no-scale SUGRA};\hfill\cr
1,\hfill \>\>\>&\mbox{beyond no-scale SUGRA.}\hfill\cr \eem
\right. \label{Fk}\eeq
The JF canonically normalized fields $\bar h_{\mu\nu}$ and
$\overline\dph$ are defined by the relations
\beq \label{Jcan1}\overline\dph=\sqrt{\frac{\vev{\bar\Omega_{\rm
H}}}{\vev{\fk}}}\dph\>\>\>\>\mbox{and}\>\>\>\> {\bar
h_{\mu\nu}\over\sqrt{2}}=
\sqrt{\vev{\fk}}\,h_{\mu\nu}+\frac{\mP\vev{\Omega_{\rm
H,\phi}}}{\sqrt{\vev{\fk}}}\eta_{\mu\nu}\dph\eeq with \beq
\bar\Omega_{\rm H}=F_{\rm K}\fk+3\mP^2 \Omega_{\rm
H,\phi}^2.\label{Jcan}\eeq\eeqs
The interaction originating from the last term in the right-hand
side of \Eref{L2} gives rise to a scattering amplitude which is
written in terms of the center-of-mass energy $E$ as follows
\beq \label{Luv} {\cal A}\sim\lf{E\over\Qef}\rg^2\>\>\>\mbox{with}
\>\>\>\Qef= {\mP\over\delta_{\cal
R}\ck^{2/n}}\frac{\vev{\bar\Omega_{\rm
H}}}{\sqrt{\vev{\fk}}}={\mP\over\delta_{\cal R}\ck^{2/n}}\lf
\frac{\vev{F_{\rm K}}}{\sqrt{2}}+3\sqrt{2}\mP^2 \vev{\Omega_{\rm
H,\phi}}^2\rg \sim\mP\eeq
(up to irrelevant numerical prefactors) since
$\vev{\fk}=1/2\ll\mP^2 \vev{\Omega_{\rm
H,\phi}}^2\simeq2^{2/n}n^2\ck^{2/n}/8$. Here $\Qef$ is identified
as the UV cut-off scale in the JF, since ${\cal A}$ remains within
the validity of the perturbation theory provided that $E<\Qef$.
Obviously, the argument above can be equally well applied to both
implementations of IG inflation in SUGRA -- see \Sref{fhi1} and
\ref{fgi} -- since the extra terms included in \Eref{Kolg} --
compared to \Eref{Kol} -- are small enough and do not generate any
problem with the perturbative unitarity.

\subsection{Einstein Frame Computation}\label{fhi3a}

Alternatively, $\Qef$ can be determined in EF, following the
systematic approach of \cref{riotto}. Note, in passing, that the
EF (canonically normalized) inflaton,
\beq\dphi=\vev{J}\dph\>\>\>\mbox{with}\>\>\>\vev{J}=\sqrt{\frac{3}{2}}\frac{n}{\vev{\xsg}}=
{\sqrt{3}\over2}\ n\sqrt[n]{2\ck} \label{dphi} \eeq
acquires mass which is given by
\beq \label{masses} \msn=\left\langle\Ve_{\rm
IG0,\se\se}\right\rangle^{1/2}= \left\langle \Ve_{\rm
IG0,\sg\sg}/J^2\right\rangle^{1/2} =\ld\mP/\sqrt{3}\ck.\eeq
Making use of \Eref{lan} we find $\msn=3\cdot10^{13}~\GeV$ for the
case of no-scale SUGRA independently of the value of $n$ -- in
accordance with the findings in \cref{gian}. Beyond no-scale
SUGRA, replacing $\ld$ in \Eref{masses} from \Eref{lang}, we find
that $\msn$ inherits from $\ld$ a mild dependence on both $n$ and
$\ksp$. E.g., for $\sg_\star=0.5\mP$, $n=2-6$ and $\ns$ in the
range of \Eref{nswmap} we find
$2.2\lesssim\msn/10^{13}~\GeV\lesssim3.8$ with the lower [upper]
value corresponding to the lower [upper] bound on $\ns$ in
\Eref{nswmap} -- see \Fref{fig2g}-({\sf\ftn a}$_1$) and ({\sf\ftn
a}$_2$).

The fact that $\dphi$ does not coincide with $\dph$ -- contrary to
the standard Higgs inflation \cite{cutoff,cutof} -- ensures that
the IG models are valid up to $\mP$. To show it, we write the EF
action ${\sf S}$ in \Eref{Saction1} along the path of \Eref{inftr}
as follows
\beqs \beq\label{S3} {\sf S}=\int d^4x \sqrt{-\what{
\mathfrak{g}}}\lf-\frac{1}{2}\mP^2 \rce +\frac12\,J^2
\dot\phi^2-\Ve_{\rm IG0}+\cdots\rg, \eeq
where the dot denotes derivation w.r.t the JF cosmic time and the
ellipsis represents terms irrelevant for our analysis. Also $J$
and $\Vhio$ are respectively given by \eqs{cannor3b}{Vhio}
[\eqs{Jg}{3Vhio}] for the model of \Sref{fhi1} [\Sref{fgi}]. For
both models, $J^2$ is accurately enough estimated by \Eref{VJe3}
-- cf. \Eref{Jg}. Expanding $J^2 \dot\phi^2$ about $\vev{\phi}$ --
see \Eref{se1} -- in terms of $\dphi$ in \Eref{dphi} we arrive at
the following result
\beq\label{exp2} J^2
\dot\phi^2=\lf1-\frac{2}{n}\sqrt{\frac{2}{3}}\frac{\dphi}{\mP}+\frac{2}{n^2}\frac{\dphi^2}{\mP^2}-
\frac{8\sqrt{2}}{3n^3\sqrt{3}}\frac{\dphi^3}{\mP^3}+\frac{20}{9n^4}\frac{\dphi^4}{\mP^4}-\cdots\rg\dot\dphi^2.\eeq
On the other hand, $\Vhio$ in \Eref{Vhio} can be expanded about
$\vev{\phi}$ as follows
\beq\Vhio=\frac{\ld^2\mP^2}{6\ck^2}\dphi^2\lf1-\sqrt{\frac{2}{3}}\lf1+{1\over
n}\rg\frac{\dphi}{\mP}+ \lf\frac{7}{18}+{1\over
n}+\frac{11}{18n^2}\rg\frac{\dphi^2}{\mP^2}-\cdots\rg\cdot\label{Vexp}\eeq\eeqs
From the expressions above, \eqs{exp2}{Vexp}, --  which reduce to
the ones presented in \cref{pallis} for $n=2$ -- we can easily
infer that $\Qef=\mP$ even for $n>2$. The same expansion is also
valid for the model of \Sref{fgi}. In any case, therefore, we
obtain $\Qef=\mP$, in agreement with our findings in \Sref{fhi3b}.

\section{Conclusions}\label{con}

In this work we showed that a wide class of IG inflationary models
can be naturally embedded in standard SUGRA. Namely, we considered
a superpotential which realize easily the IG idea and can be
uniquely determined by imposing two global symmetries -- a
continuous $R$ and a discrete $\mathbb{Z}_n$  symmetry -- in
conjunction with the requirement that inflation has to occur for
\sub\ values of the inflaton. On the other hand, we adopted two
forms of \Ka s, one corresponding to the \Km\ $SU(2,1)/SU(2)\times
U(1)_R\times\mathbb{Z}_n$, inspired by no-scale SUGRA, and one
more generic. In both cases, the tachyonic instability, occurring
along the direction of the accompanying non-inflaton field, can be
remedied by considering terms up to the fourth order in the \Ka.
Thanks to the underlying symmetries the inflaton, $\phi$ appears
predominantly as $\phi^n$ in both the super- and \Ka s.

In the case of no-scale SUGRA,  the inflaton is not mixed with the
accompanying non-inflaton field in \Ka. As a consequence, the
model predicts results identical to the non-SUSY case
independently of the exponent $n$. In particular, we found
$\ns\simeq0.963$, $\as\simeq-0.00068$ and $r\simeq0.0038$, which
are in excellent agreement with the current data, and
$\msn=3\cdot10^{13}~\GeV$. Beyond no-scale SUGRA, all the possible
terms up to the forth order in powers of the various fields are
included in the \Ka. In this case, we can achieve $\ns$ precisely
equal to its central observationally favored value, mildly tuning
the coefficient $\ksp$. Furthermore, a weak dependance of the
results on $n$ arises with the lower $n$'s being more favored,
since the required tuning on $\ksp$ is softer. In both cases a
$n$-dependent lower bound on $\ck$ assists us to obtain inflation
for \sub\ values of the inflaton, stabilizing thereby our proposal
against possible corrections from higher order terms in $\fk$.
Furthermore we showed that the one-loop radiative corrections
remain subdominant during inflation and the corresponding
effective theory is trustable up to $\mP$. Indeed, these models
possess a built-in solution into long-standing naturalness problem
\cite{cutoff,riotto} which plagued similar models. As demonstrated
both in the EF and the JF, this solution relies on the dynamical
generation of $\mP$ at the vacuum of the theory.

As a bottom line we could say that although no-scale SUGRA has
been initially coined as a solution to the problem of SUSY
breaking \cite{noscale,eno9} ensuring a vanishing cosmological
constant, it is by now recognized -- see also
\cite{eno7,zavalos,pallis} -- that it provides a flexible
framework for inflationary model building. In fact, no-scale SUGRA
is tailor-made for IG (and nonminimal, in general) inflation since
the predictive power of this inflationary model in more generic
SUGRA incarnations is lost.

\subsubsection*{\large\bfseries\scshape Note Added}

When this work was under completion, the {\small\sc Bicep2}
experiment \cite{gws} announced the detection of B-mode
polarization in the cosmic microwave background radiation at large
angular scales. If this mode is attributed to the primordial
gravity waves predicted by inflation, it implies \cite{gws}
$r=0.16^{+0.06}_{-0.05}$ -- after subtraction of a dust model.
Combining this result with \sEref{nswmap}{c} we find -- cf.
\cref{rcom} -- a simultaneously compatible region $0.06\lesssim
r\lesssim0.135$ (at $95\%$ c.l.) which, obviously, is not
fulfilled by the models presented here, since the predicted $r$
lies one order of magnitude lower -- see \Eref{res} and comments
below \Eref{resg3}. However, it is still premature to exclude any
inflationary model with $r$ lower than the above limit since the
current data are subject to considerable foreground uncertainty --
see e.g. \cref{gws1,gws2}.

\begin{acknowledgement}

This research was supported by the Generalitat Valenciana under
contract PROMETEOII/2013/017.

\end{acknowledgement}

\newpage

\rhead[\fancyplain{}{ \bf \thepage}]{\fancyplain{}{\sl IG
Inflation in no-Scale SUGRA \& Beyond}} \lhead[\fancyplain{}{\sl
\leftmark}]{\fancyplain{}{\bf \thepage}} \cfoot{}

\end{document}